\documentclass[reprint, double column, aps, prb, superscriptaddress, showkeys]{revtex4-2}

\usepackage{graphicx,epsfig}
\usepackage{amssymb}
\usepackage{amsmath}
\usepackage{bm}
\usepackage{booktabs}
\usepackage{textcomp}
\usepackage{color}
\usepackage{multirow}
\usepackage{braket}

\usepackage{soul}

\usepackage[bookmarksnumbered,bookmarksopen]{hyperref}
\hypersetup{
   colorlinks=true, linkcolor=blue      
}

\begin{document}
\setcounter{page}{1}

\title[]{Competing Many-Body Phases at Small Fillings in Ultrahigh-Quality GaAs 2D Hole Systems: Role of Landau Level Mixing}
\author{Chengyu \surname{Wang}}
\author{A. \surname{Gupta}}
\author{S. K. \surname{Singh}}
\author{L. N. \surname{Pfeiffer}}
\author{K. W. \surname{Baldwin}}
\affiliation{Department of Electrical and Computer Engineering, Princeton University, Princeton, New Jersey 08544, USA}
\author{R. \surname{Winkler}}
\affiliation{Department of Physics, Northern Illinois University, DeKalb, Illinois 60115, USA}
\author{M. \surname{Shayegan}}
\affiliation{Department of Electrical and Computer Engineering, Princeton University, Princeton, New Jersey 08544, USA}
\date{\today}

\begin{abstract}
The fractional quantum Hall state (FQHS), an incompressible liquid state hosting anyonic excitations with fractional charge and statistics, represents a compelling many-body phase observed in clean two-dimensional (2D) carrier systems. The expected non-Abelian nature of the FQHSs at even-denominator Landau level (LL) fillings has particularly sparked considerable recent interest. At sufficiently small fillings, another exotic phase, namely a quantum Wigner crystal (WC) state, dominates. Here we report magneto-transport measurements in an ultrahigh-quality GaAs 2D \textit{hole} system where the large hole effective mass leads to a significant LL mixing (LLM) even at very high magnetic fields and affects the many-body states at very small fillings. We observe numerous developing FQHSs at both even- and odd-denominator fillings, deep in the insulating regime at $\nu \lesssim 1/3$ where WC states dominate. The FQHSs we observe at odd-denominator fillings on the flanks of $\nu=1/4$ and 1/6 are consistent with the Abelian Jain sequence of four-flux and six-flux composite fermions, while the ones at even-denominator fillings $\nu=1/4$ and 1/6 are likely non-Abelian states emerging from the pairing of these quasiparticles induced by severe LLM. Our results demonstrate that the competition between the FQHSs and WC phases is close at very small fillings even in the presence of severe LLM. We also measure activation energies of WC states near $\nu=1/6$, and find that they are substantially larger than what has been reported for ultrahigh-quality GaAs 2D electrons. A moderate LLM is believed to lower the activation energy associated to the formation of WC intrinsic defects. The surprisingly large activation energy for our 2DHS with significant LLM is therefore puzzling, and may suggest a different type of intrinsic WC defect compared to that in 2D electrons.
\end{abstract}

\maketitle
\section{\label{sec:level1}Introduction}
Low-disorder, two-dimensional (2D) charged carrier systems subjected to large perpendicular magnetic fields at low temperatures are consummate platforms for the study of many-body quantum phenomena thanks to the quenched thermal and kinetic energies. When the lowest-energy Landau level (LL) is partially occupied, the dominant electron-electron Coulomb interaction leads to the emergence of a variety of strongly-correlated, many-body quantum phases \cite{Tsui.PRL.1982, Shayegan.review.2006, Jain.Book.2007, Halperin.Jain.Book.2020}. At sufficiently small LL fillings ($\nu$), a disorder-free (ideal) 2D electron system (2DES) is expected to form an ordered array, the so-called quantum Wigner crystal (WC) \cite{Wigner.PR.1934}, when the long-range part of Coulomb interaction is dominant. In realistic (non-ideal) samples, the many-body electronic crystal breaks into domains pinned by ubiquitous disorder, exhibiting a highly-insulating behavior \cite{Shayegan.NatRevPhys.2022}. Evidence for pinned WC phases has been reported in the lowest LL of different 2D systems, including GaAs 2D electrons \cite{Andrei.PRL.1988, Jiang.PRL.1990, Goldman.PRL.1990, Chen.NatPhys.2006, Deng.PRL.2016, Deng.PRL.2019} and holes \cite{Santos.PRL.1992, Santos.PRB.1992, Ma.PRL.2020}, AlAs \cite{Kevin.PRR.2021, Singh.PRB.2024}, ZnO \cite{Maryenko.NC.2018}, and bilayer graphene \cite{Tsui.Nature.2024}. At certain rational fractional fillings, the fractional quantum Hall state (FQHS), a strongly-correlated, incompressible liquid with nontrivial topological properties, manifests itself as the ground state thanks to the strong short-range electron-electron correlation \cite{Tsui.PRL.1982, Shayegan.review.2006, Jain.Book.2007, Halperin.Jain.Book.2020}. The fractional charge and anyonic statistics \cite{Nakamura.Natphys.2020, Bartolomei.science.2020, Feldman.RPP.2021}, as well as the potentially non-Abelian nature of the quasiparticle excitations of even-denominator FQHSs \cite{Nayak.RMP.2008, Banerjee.Nature.2018, Willett.PRX.2023} have further sparked intense research interest in recent years.

A partially filled lowest LL with small fillings is of particular interest because it serves as the battleground for fundamental liquid and solid phases of interacting carriers. Experiments on ultrahigh-quality GaAs 2DESs revealed evidence for developing FQHSs at specific, extremely small fillings, e.g., at $\nu=$ 1/7 \cite{Goldman.PRL.1988, Pan.PRL.2002, Chung.PRL.2022, Wang.Preprint.2024}, although insulating phases become dominant in filling ranges 1/5 $<\nu<$ 2/9 and $\nu<$ 1/5 \cite{Jiang.PRL.1990, Goldman.PRL.1990, Chen.NatPhys.2006, Deng.PRL.2016, Deng.PRL.2019}. These observations signal the intricate competition between strongly correlated liquid and solid states, namely, FQHSs and a quantum WC.

\begin{figure*}[t!]
  \begin{center}
    \psfig{file=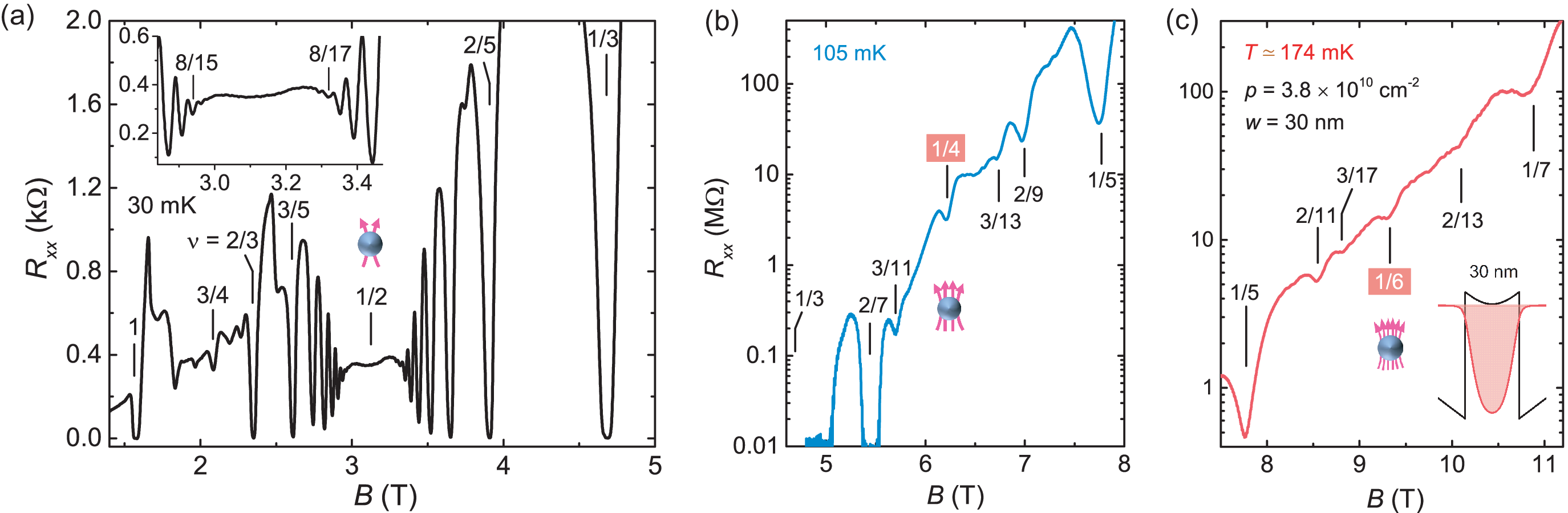, width=1\textwidth}
  \end{center}
  \caption{\label{summary}
    Magneto-transport data for 2D holes confined to a 30-nm-wide GaAs quantum well in different filling factor ranges: (a) $1>\nu>1/3$, (b) $1/3>\nu>1/5$, and (c) $1/5>\nu>1/7$. Traces in (a) are shown in a linear scale, with k$\Omega$ as units of $R_{xx}$. Traces in (b) and (c) are shown with a \textit{log} scale for $R_{xx}$ and in units of M$\Omega$. The $B$ positions of several fractional filling factors are marked with black vertical lines. Inset in (a): An enlarged version of the $R_{xx}$ vs $B$ near $\nu=$1/2. Inset in (c): The self-consistently calculated hole charge distribution (red) and potential (black).
    }
  \label{fig:summary}
\end{figure*}

In GaAs 2D \textit{hole} systems (2DHSs), the larger effective mass ($\sim 0.5m_0$ \cite{Zhu.SSC.2007}) compared to electrons (0.067$m_0$) suppresses the LL separation at high $B$, leading to a significant mixing of higher LLs into the many-body ground states of the lowest LL \cite{Footnote.mixing, Winkler.Book.2003}. Such mixing is quantified by the LL mixing (LLM) parameter $\kappa$, defined as the ratio of Coulomb energy $E_{\text{Coul}}$ and cyclotron energy $E_{\text{cyc}}$. LLM explicitly breaks the particle-hole symmetry within a LL \cite{Liu.PRB.2014}. More importantly, it can modify the electron-electron interaction and play an important role in the stabilization of different correlated states in both lowest and excited LLs \cite{Santos.PRL.1992, Santos.PRB.1992, Peterson.PRL.2014, Kevin.PRR.2021, Singh.PRB.2024, Rezayi.PRL.2017, Zhao.PRL.2018, Ma.PRL.2020, Hossain.NatPhys.2021, Wang.PRL.2022, Wang.PRL.2023.1/4, Wang.PNAS.2023, Zhao.PRL.2023, Das.PRL.2023}. For example, moderate LLM can affect the candidate non-Abelian FQHS at $\nu=$ 5/2 \cite{Peterson.PRL.2014, Rezayi.PRL.2017, Das.PRL.2023}. In the lowest LL of dilute GaAs 2DHSs, severe LLM leads to the onset of insulating phases near $\nu\sim1/3$ \cite{Santos.PRL.1992, Santos.PRB.1992, Ma.PRL.2020} rather than $\nu\sim1/5$. This is because LLM softens the short-range part of Coulomb interaction, thereby favoring the formation of WC states over liquid states. Similar phenomena were also reported in other materials such as AlAs \cite{Kevin.PRR.2021, Singh.PRB.2024}, ZnO \cite{Maryenko.NC.2018}, and bilayer graphene \cite{Tsui.Nature.2024}. With recent advancements in the quality of GaAs 2DHSs \cite{Chung.PRM.2022, Gupta.PRM.2024}, many exotic FQHSs, likely emerging from composite fermion (CF) interaction and pairing induced by LLM, were observed at \textit{even-denominator} fillings in the lowest LL \cite{Wang.PRL.2022, Wang.PNAS.2023, Wang.PRL.2023.1/4}.

Here we report the observation of numerous developing FQHSs in the lowest LL of ultrahigh-quality GaAs 2DHSs at \textit{even-denominator} fillings $\nu=$ 1/6 and 1/4, and at odd-denominator fillings that follow the Jain sequence of four-flux and six-flux CFs ($^4$CFs and $^6$CFs). These states closely compete with pinned WC states which are dominant at $\nu<$ 1/3. Our observations highlight the unexpected robustness of Jain FQHSs at extremely small fillings in the presence of severe LLM. They also suggest that, with sufficiently large LLM, the $^6$CFs at $\nu=$ 1/6 may undergo a pairing instability and condense into a non-Abelian FQHS, similar to what was recently reported at $\nu=$ 1/4 \cite{Wang.PRL.2023.1/4, Zhao.PRL.2023}.

\section{\label{sec:level1}Experimental details}
We studied an ultrahigh-quality 2DHS confined to a 30-nm-wide GaAs quantum well (QW) grown on a GaAs (001) substrate by molecular beam epitaxy. The 2DHS has a density of 3.8$\times$10$^{10}$ cm$^{-2}$ and a low-temperature (0.03 K) record-high mobility of 18$\times$10$^{6}$ cm$^2$/Vs. The exceptionally high mobility was achieved by optimizing the growth chamber vacuum integrity and the purity of the source materials \cite{Chung.NatMater.2021}, as well as an optimized stepped-barrier design \cite{Chung.PRM.2022, Gupta.PRM.2024}. We performed our experiments on 4$\times$4 mm$^2$ van der Pauw geometry samples cleaved from 2-inch GaAs wafers, and with alloyed In:Zn contacts at the four corners and side midpoints. The samples were cooled in a dilution refrigerator and magneto-transport measurements were performed using low-frequency, lock-in amplifier techniques. A small excitation current of $I<$ 0.5 nA was used for measuring the resistance of insulating phases to avoid self-heating and nonlinear $I$-$V$ effect.

\section{\label{sec:level1}Magneto-transport data}

\subsection{\label{sec:level2}Developing fractional quantum Hall states at small $\nu$}

Figure \ref{fig:summary}(c) highlights our main findings: We observe numerous developing FQHSs on the flanks of $\nu=$ 1/6 at odd-denominator fillings $\nu=$ 1/7, 2/13, and 1/5, 2/11, 3/17, evinced by sharp $R_{xx}$ minima superimposed on a highly-insulating background. These states can be interpreted as Jain-sequence states (i.e., integer QHSs) of $^6$CFs [at $\nu=n/(6n\pm1)$, $n=1,2,3$]. Remarkably, our data also exhibit a reasonably sharp $R_{xx}$ minimum at $\nu=$ 1/6, suggesting the emergence of an \textit{even-denominator} FQHS, likely stabilized by the residual interaction and pairing of $^6$CFs. The observation of developing, even- and odd-denominator FQHSs at such small fillings is surprising because our 2DHS experiences severe LLM ($\kappa\simeq6$ near $\nu=$ 1/6), and large $\kappa$ generally favors WC states over FQH liquid states at small $\nu$ \cite{Zhao.PRL.2018, Ma.PRL.2020, Kevin.PRR.2021}.

\begin{figure*}[t!]
  \begin{center}
    \psfig{file=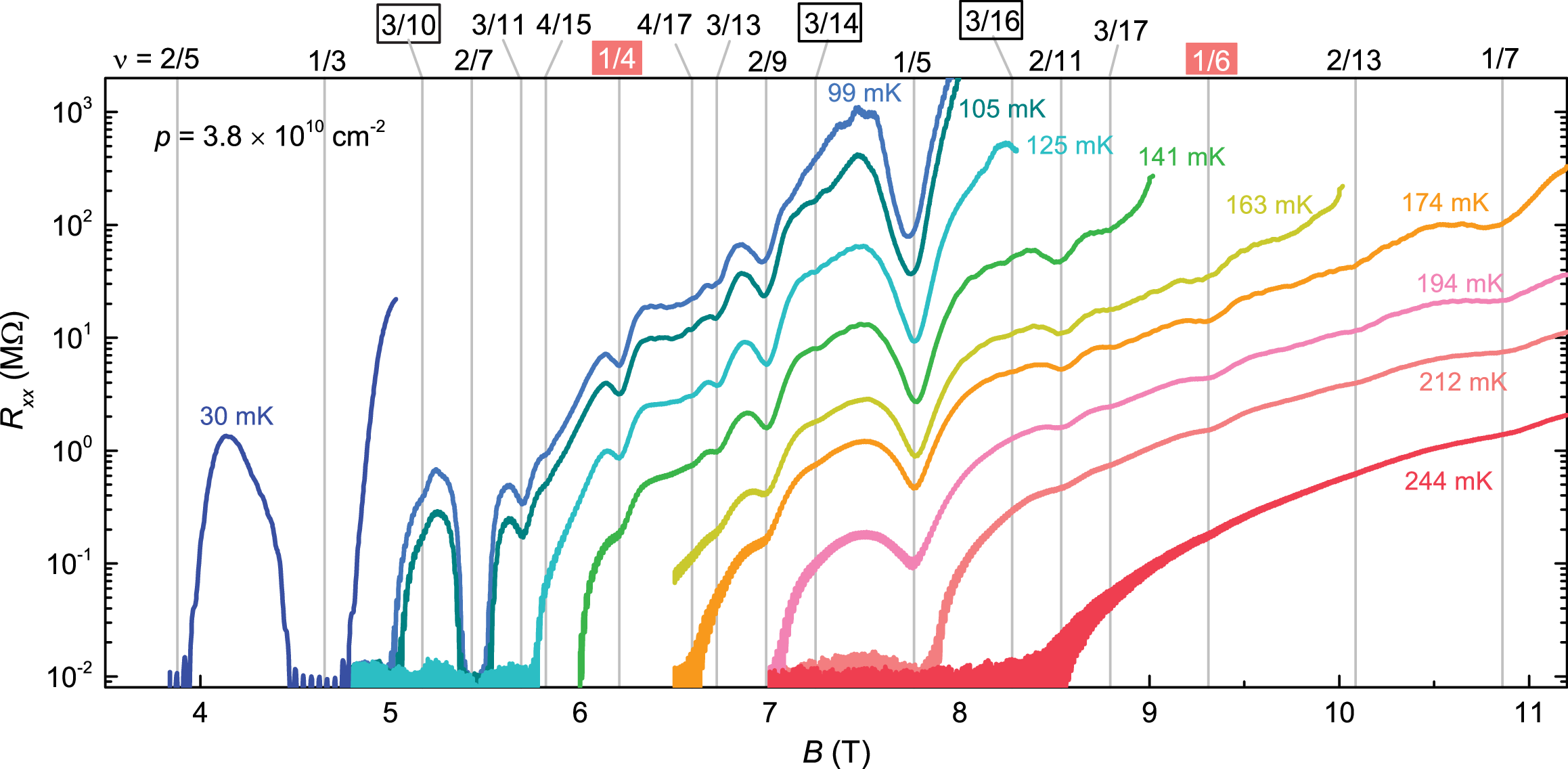, width=1\textwidth}
  \end{center}
  \caption{\label{Tdep} 
     $R_{xx}$ vs $B$ traces for $\nu\leq2/5$ taken at different temperatures. The vertical lines mark the $B$ positions of fractional filling factors as indicated on the top axis.
}
  \label{fig:Tdep}
\end{figure*}

At lower magnetic fields, we observe a developing even-denominator FQHS at $\nu=1/4$, flanked by numerous FQHSs at odd-denominator fillings $\nu=$ 1/5, 2/9, 3/13, and 1/3, 2/7, 3/11 [Fig. \ref{fig:summary}(b)]. These observations are consistent with what was reported in ultrahigh-quality 2DHSs very recently \cite{Wang.PRL.2023.1/4}, and the $\nu=1/4$ FQHS is believed to emerge from the $^4$CF pairing induced by substantial LLM \cite{Zhao.PRL.2023}. The odd-denominator FQHSs on the flanks of $\nu=1/4$ can be interpreted as Jain-sequence states of $^4$CF [$\nu=n/(4n\pm1)$, $n=1,2,3$]. It is noteworthy that the data shown in Figs. \ref{fig:summary}(b) and \ref{fig:summary}(c) were taken at elevated temperatures of $T\simeq$105 mK and $\simeq$174 mK, respectively, so that the 2DHS was less insulating (less resistive) and $R_{xx}$ could be measured reliably. Figure \ref{fig:summary}(a) shows a trace measured at the base temperature $T\simeq$30 mK for $1>\nu>1/3$. Near $\nu=$ 1/2, the sample exhibits a smooth, shallow minimum, consistent with a compressible CF Fermi sea ground state at $\nu=$ 1/2 \cite{Willett.PRL.1993}. The $\nu=$ 1/2 minimun is flanked by numerous Jain-sequence FQHSs of two-flux CFs ($^2$CFs) at $\nu=n/(2n\pm1)$ ($n=1, 2, ..., 8$). We also observe a developing, exotic FQHS at $\nu=$ 3/4 which can be interpreted as an even-denominator FQHS of $^2$CFs arising from the CF-CF interaction and pairing in the half-filled, excited CF LL \cite{Wang.PRL.2022, Wang.PNAS.2023}.

\begin{figure}[t!]
  \begin{center}
    \psfig{file=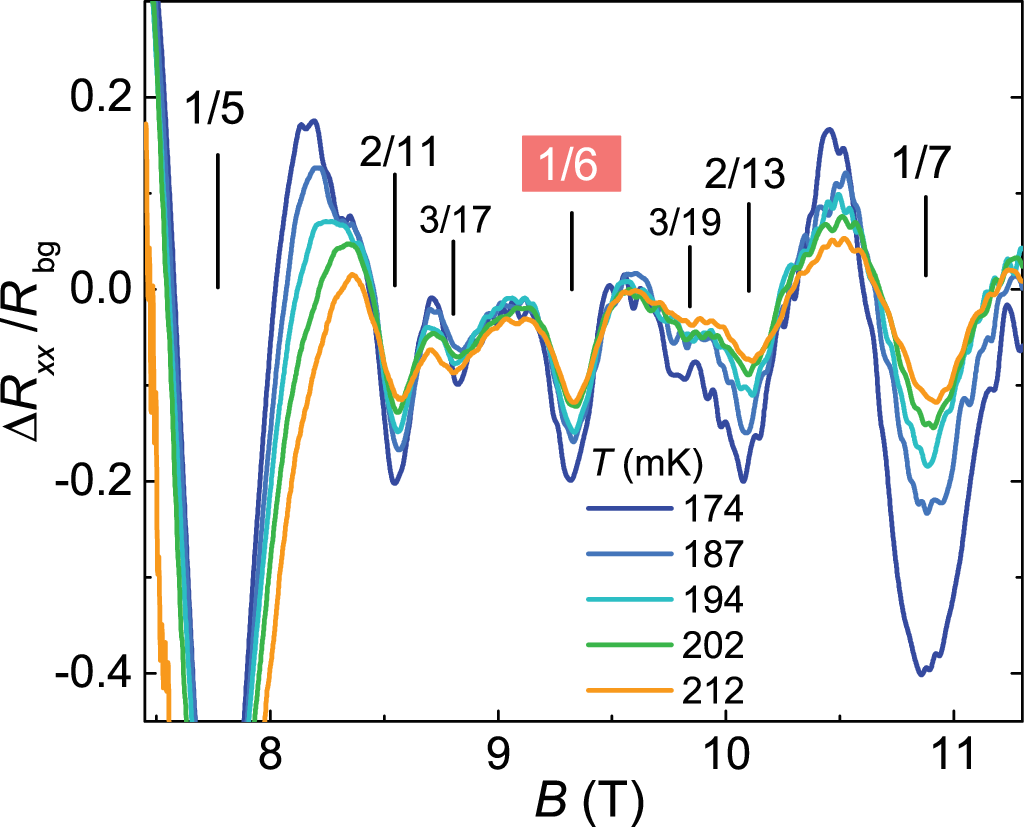, width=0.4
    \textwidth}
  \end{center}
  \caption{\label{deltaR} 
   $\Delta R_{xx}/R_{bg}$ vs $B$ traces at different temperatures, where $\Delta R_{xx}$  represents the resistance after subtracting a smooth, increasing background $R_{bg}$. This background is estimated using a Savitzky-Golay filter with a 1 T window applied to $R_{xx}$ vs $B$. We observe sharp $R_{xx}$ minima at $\nu=$ 1/5, 2/11, 3/17, 2/13, and 1/7. The strengths of these minima weaken as $\nu$ is approaching 1/6, and also weaken with increasing $T$, consistent with what is expected for Jain-sequence FQHSs. Additionally, we observe a sharp minimum at $\nu=1/6$, closely resembling those observed at Jain-sequence fillings and suggesting a developing FQHS at $\nu=1/6$. }
  \label{fig:deltaR}
\end{figure}

In Fig. \ref{fig:Tdep}, we present $R_{xx}$ vs $B$ traces for $\nu<$ 2/5 taken at different temperatures. FQHSs are fully developed at $\nu=$ 2/5, 1/3, and 2/7, evinced by vanishing $R_{xx}$ and flat Hall plateaus quantized at $R_{xy}=h/\nu e^2$; see Fig. \ref{fig:Hall} in Appendix A for Hall data. Between these fully developed FQHSs, i.e., in filling ranges $2/5>\nu>1/3$ and 1/3 $>\nu>$ 2/7, as well as at higher $B$ ($\nu<$ 2/7), the data clearly reveal an insulating behavior: The $R_{xx}$ values are extremely high ($>1$ M$\Omega$) at low temperatures and sharply decrease with increasing temperature. Such insulating phases were first reported in GaAs 2DESs for $\nu<$ 1/5 and 2/9 $>\nu>$ 1/5 \cite{Andrei.PRL.1988, Jiang.PRL.1990, Goldman.PRL.1990, Chen.NatPhys.2006, Deng.PRL.2016, Deng.PRL.2019}, and are believed to signal the formation of pinned WC phases. In systems with large LLM like our GaAs 2DHS, experiments \cite{Santos.PRL.1992, Santos.PRB.1992, Ma.PRL.2020, Kevin.PRR.2021, Singh.PRB.2024, Maryenko.NC.2018, Tsui.Nature.2024} and theoretical calculations \cite{Zhao.PRL.2018} show that the onset of insulating phases moves to larger fillings, near $\nu=$ 1/3, consistent with our data in Fig. \ref{fig:Tdep}. 

\begin{figure}[t!]
  \begin{center}
    \psfig{file=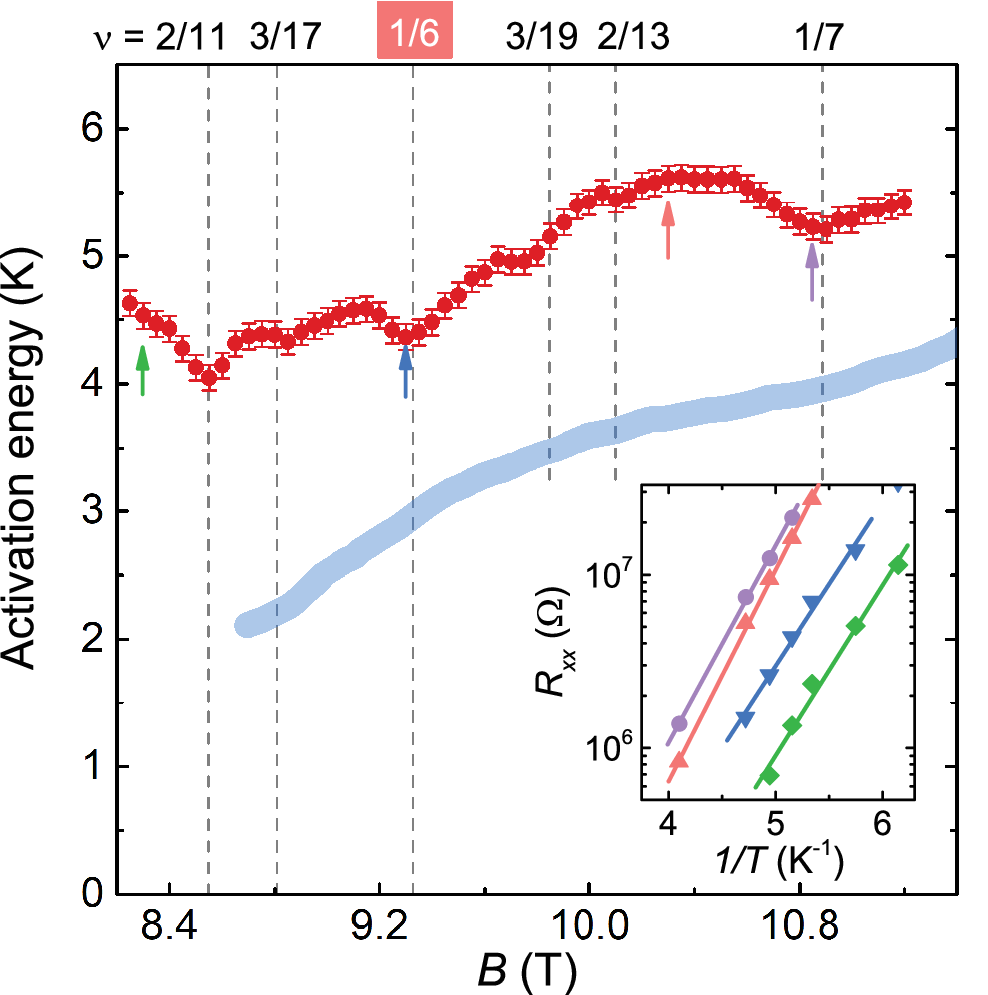, width=0.48 \textwidth}
  \end{center}
  \caption{\label{activation} 
   Activation energy vs $B$ for $R_{xx}$ data shown in Fig. \ref{fig:Tdep} at extremely small fillings $\nu<$ 1/5. The blue curve represents theoretical values of the CF WC defect formation energy reported in Ref. \cite{Archer.PRB.2014}. Inset shows $R_{xx}$ vs $1/T$ Arrhenius plots at various $B$, each color coded according to the $B$ values marked by the arrows in the main figure.
  }
  \label{fig:activation}
\end{figure}

\begin{figure*}[t!]
  \begin{center}
    \psfig{file=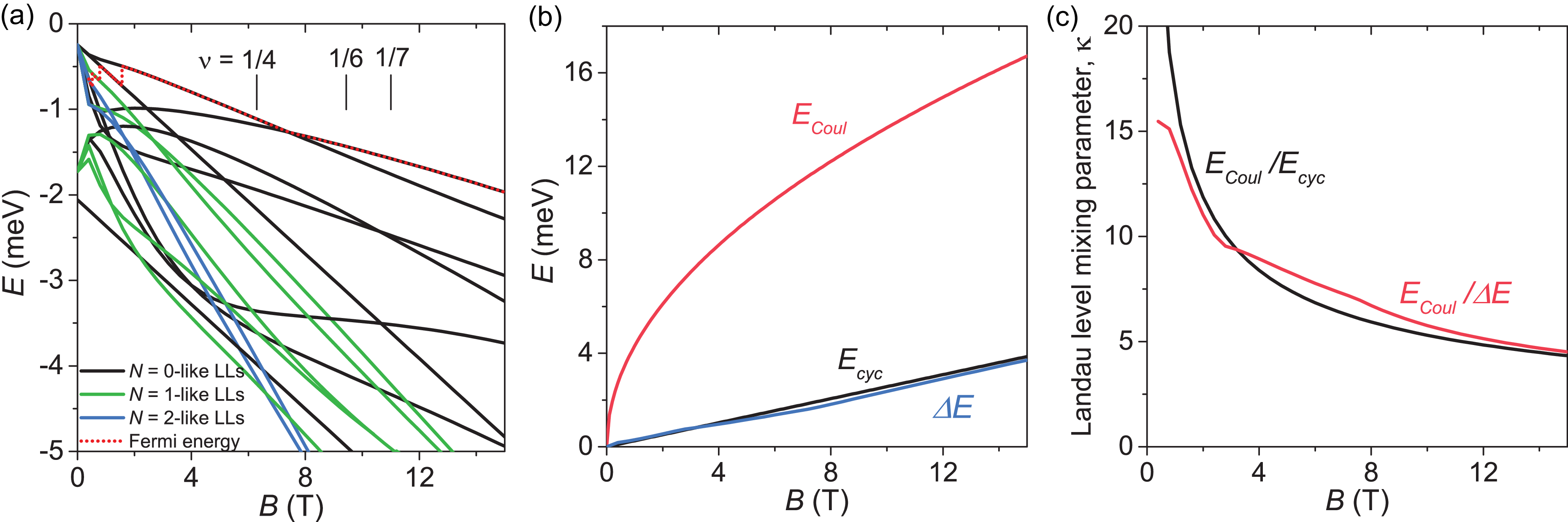, width=1 \textwidth}
  \end{center}
  \caption{\label{LLanalysis} 
   (a) Calculated energy $E$ vs $B$ Landau level diagram. Different LLs are color-coded according to their dominant orbital component in the field range 4 $<B<$ 12 T where competing FQHSs and WC states are observed. The red dotted line traces the Fermi energy $E_F$. LLs with energy much lower than $E_F$ are not shown for simplicity. (b) Coulomb energy $E_{\text{Coul}}$, cyclotron energy $E_{\text{cyc}}$, and the separation $\Delta E$ between the topmost $N=0$-like and $N=1$-like LLs for our 2DHS as a function of $B$. (c) Landau level mixing parameters according to two different definitions.
  }
  \label{fig:LLanalysis}
\end{figure*}

Note in Fig. \ref{fig:Tdep} that the $R_{xx}$ minima at $\nu=$ 1/6 and 1/4 are observed in a reasonably wide range of temperatures (163 $\sim$ 194 mK for $\nu=$ 1/6, and 99 $\sim$ 125 mK for $\nu=$ 1/4). With further increase in $T$, these minima turn into inflection points and eventually disappear. Such evolution of $R_{xx}$ minima at $\nu=$ 1/4 and 1/6 with increasing $T$ is very similar to what is observed at Jain-sequence odd-denominator fillings flanking $\nu=1/4$ and 1/6, supporting that they are indeed developing FQHSs. We also note that between the $n=1$ and $n=2$ Jain-sequence states, instead of a smooth $R_{xx}$ dome, our sample exhibits inflection points at $\nu=$ 3/10, 3/14, and 3/16, also evinced by peaks in $d^2R_{xx}/dB^2$ vs $B$; see Fig. \ref{fig:derivative} in Appendix B for more details. These features could be precursors of developing even-denominator FQHSs of $^4$CFs and $^6$CFs, analogous to the exotic $\nu=$ 3/4 FQHS \cite{Wang.PRL.2022, Wang.PNAS.2023}.

In order to more clearly reveal the $R_{xx}$ minima and inflection points riding on the insulating phases for $\nu\lesssim 1/5$, Fig. \ref{fig:deltaR} presents $\Delta R_{xx}/R_{bg}$ vs $B$ traces at various temperatures. Here, $\Delta R_{xx}$ represents the resistance after subtracting a smooth, increasing background $R_{bg}$; see Fig. \ref{fig:background} and Appendix B for details. We observe sharp $\Delta R_{xx}/R_{bg}$ minima at $\nu=$ 1/5, 2/11, 3/17, 2/13, and 1/7. The strength of these minima weakens as $\nu$ is approaching 1/6, and also weakens with increasing $T$, consistent with what is expected for Jain-sequence FQHSs. Additionally, we observe a sharp minimum at $\nu=1/6$, closely resembling those observed at Jain-sequence fillings and suggesting a developing FQHS at $\nu=1/6$.

\subsection{\label{sec:level2}Activation energy of Wigner crystal states}

Next, we study the WC states that compete with FQHSs at extremely small fillings. In Fig. \ref{fig:activation}, we present activation energy ($E_A$) data deduced from the relation $R_{xx}\propto e^{E_A/2kT}$ as a function of $B$. The inset in Fig. \ref{fig:activation} shows representative $R_{xx}$ vs $1/T$ Arrhenius plots and the linear fits whose slopes give the value of $E_A$ at a given $B$. The magnitude of $E_A$ is associated with the defect formation energy of a WC. Numerical calculations indicate that, at small fillings ($\nu<$ 1/5), a WC formed by $^4$CFs has lower energy than the conventional Hartree-Fock electron crystal \cite{Archer.PRL.2013}, and that the lowest-energy \textit{intrinsic} defect in this $^4$CF WC is a hyper-correlated quantum bubble defect \cite{Archer.PRB.2014}. Recent measurements \cite{Madathil.PRL.2024} on ultrahigh-quality 2DESs indeed show $E_A$ values closely matching with that expected for an ideal $^4$CF WC with no disorder, consistent with the fact that the WC domains in these samples are estimated to contain $\simeq$ 1000 electrons \cite{Madathil.PRL.2023}. Experiments on modest quality GaAs 2DESs \cite{Willett.PRB.1988, Jiang.PRB.1991} and ultrahigh-quality 2DESs in the dilute limit \cite{Madathil.PRL.2024} show somewhat smaller values of $E_A$ compared to calculations. These discrepancies were attributed to the effects of disorder and LLM which are generally believed to lower $E_A$. Surprisingly, our 2DHS exhibits $E_A$ significantly \textit{larger} than what was reported for 2DESs at a similar density of 4.4$\times$10$^{10}$ cm$^{-2}$ \cite{Madathil.PRL.2024}, even though the LLM parameter $\kappa$ for our 2DHS ($\kappa\simeq$ 5) is about seven times larger than $\kappa$ for the 2DES ($\kappa\simeq$ 0.7). The $E_A$ values for our 2DHS are even larger than the calculated values reported in Ref. \cite{Archer.PRB.2014}; see the blue curve in Fig. \ref{fig:activation}. The anomalously large measured $E_A$ suggests that the nature of the WC and/or the lowest-energy intrinsic defect of the WC in our ultrahigh-quality 2DHS might be of a different type.

\section{\label{sec:level1}Discussion}

\subsection{\label{sec:level2}Estimate of Landau level mixing parameter $\kappa$ for GaAs 2DHSs}

It is noteworthy that, unlike GaAs 2DESs whose LLs are linear in energy as a function of $B$, GaAs 2DHSs have highly nonlinear LLs because of the heavy-hole light-hole coupling \cite{Footnote.mixing, Winkler.Book.2003}; see e.g. Fig. \ref{fig:LLanalysis}(a). The complex LL diagram exhibits numerous LL crossings that can be tuned to coincide with the Fermi energy ($E_F$), giving rise to exotic interaction phenomena \cite{Liu.PRB.2014.1/2, Liu.PRB.2016, Lupatini.PRL.2020, Ma.PRL.2022, Wang.PRL.2023}. On the other hand, this nonlinearity make a quantitative evaluation of the cyclotron energy and LLM parameter $\kappa$ more challenging \cite{Ma.PRL.2022, Wang.PRL.2023.1/4}. In addition, the heavy-hole light-hole and spin-orbit couplings in holes render hole LLs multi-component, making the orbital quantum number $N$ not a good quantum number for a hole LL \cite{Winkler.Book.2003, Broido.PRB.1985}. This adds complexity to the definition of cyclotron energy and $\kappa$. In the remainder of this article, we present LL calculations for our 2DHS to shed light on the estimation and effect of $\kappa$ in the extreme quantum limit. 

We calculated the $E$ vs $B$ LL diagram for our 2DHS, using the multiband envelope function approximation and the 8$\times$8 Kane Hamiltonian \cite{Winkler.Book.2003}. For simplicity, in Fig. \ref{fig:LLanalysis}(a), we only show the topmost LLs which are most relevant to our study. We expand these LLs in a basis of four spinors, each of which can be described by a spin quantum number ($s_z=\pm$1/2 or $\pm$3/2) and an orbital quantum number ($N=0, 1, 2...$); see Appendix E for detailed results of our calculations. The LLs in Fig. \ref{fig:LLanalysis}(a) are color-coded based on $N$ of the leading spinor component in the range $4<B<12$ T. Figure \ref{fig:LLanalysis}(b) shows the different energy scales in the extreme quantum limit: (i) Coulomb energy $E_{\text{Coul}}=e^2/(4\pi\epsilon l_B)\propto\sqrt{B}$, where $l_B=\sqrt{\hbar/eB}$ is the magnetic length; (ii) cyclotron energy $E_{\text{cyc}}=\hbar eB/m^*_0\propto B$, where $m_0=0.45$ (in units of free electron mass) is the calculated band mass at $B=0$; and (iii) $\Delta E$, defined as the separation between the topmost $N=0$-like LL and $N=1$-like LL in Fig. \ref{fig:LLanalysis}(a) \cite{Footnote.deltaE}. 

We find that in Fig. \ref{fig:LLanalysis}(b), $E_{\text{cyc}}$ and $\Delta E$ are close throughout the range of $B$. Such an agreement is surprising because the former is from a $B=0$ energy band calculation while the latter is extracted from the LL calculations in large $B$ where the LLs are complex and nonlinear. It is worth noting that the simple picture of $E_{\text{cyc}}=\hbar eB/m^*_0$ has been used to estimate $\kappa$ for GaAs 2DHSs in the past \cite{Santos.PRB.1992, Ma.PRL.2020}. A reasonably good agreement on $\kappa$ vs $\nu$ WC quantum melting phase diagram near $\nu=1/3$ was found between experimental results on GaAs 2DHSs \cite{Ma.PRL.2020} and calculations \cite{Zhao.PRL.2018}. As we will discuss later, our data points are also consistent with the calculated phase diagram; see Fig. \ref{fig:phase}. The $E_{\text{cyc}}\simeq \Delta E$ we find in Fig. \ref{fig:LLanalysis}(b) may explain these agreements.

\subsection{\label{sec:level2}Competition between FQHSs and WC states: Role of LLM}

The fact that $E_{\text{Coul}}$ is significantly larger than $E_{\text{cyc}}$ and $\Delta E$ highlights the importance of LLM. The ratios $E_{\text{Coul}}/E_{\text{cyc}}$ and $E_{\text{Coul}}/\Delta E$ offer two estimates for $\kappa$ in the extreme quantum limit [Fig. \ref{fig:LLanalysis}(c)]. They both yield $9\gtrsim\kappa\gtrsim5$ for $4<B<12$ T \cite{Footnote.agreement}. The large $\kappa\simeq5$ near $\nu=$ 1/7 in our 2DHS is about six to ten times larger than $\kappa$ of GaAs 2DESs where a developing $\nu=$ 1/7 FQHS was reported \cite{Goldman.PRL.1988, Pan.PRL.2002, Chung.PRL.2022, Wang.Preprint.2024}. 

\begin{figure*}[t]
  \begin{center}
    \psfig{file=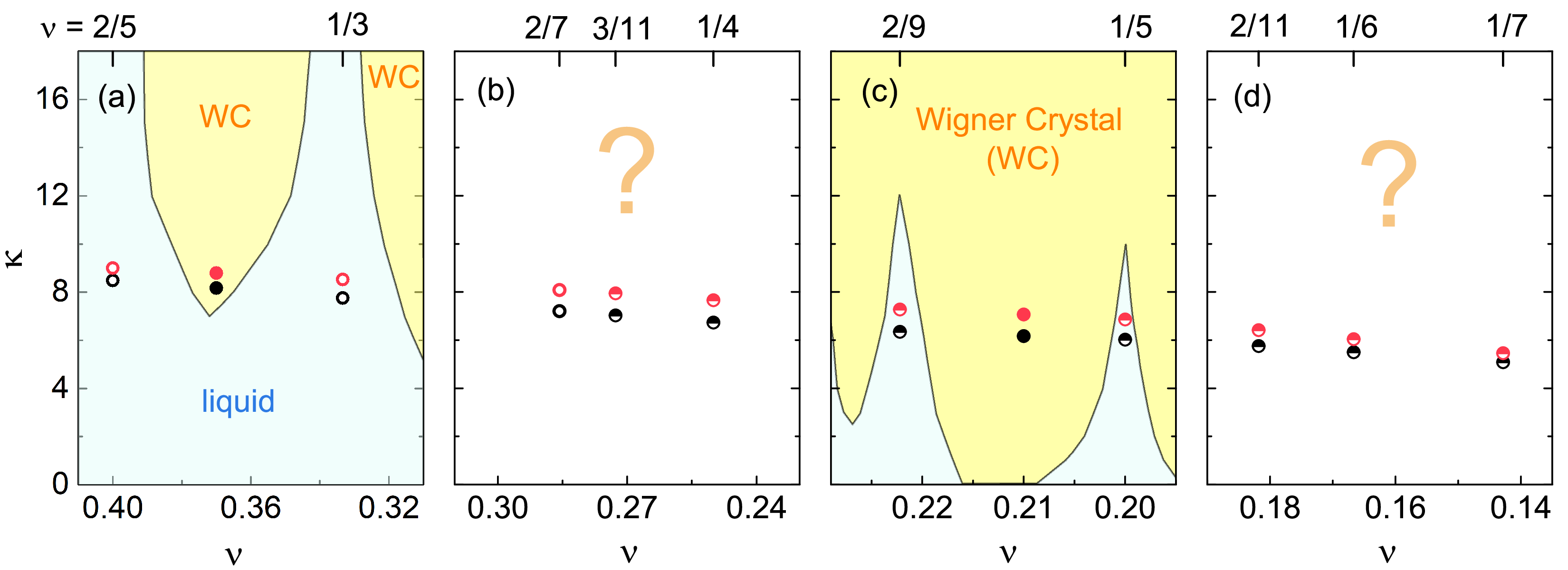, width=0.9 \textwidth}
  \end{center}
  \caption{\label{fig:phase}WC quantum melting phase diagrams, with LLM parameter $\kappa$ and $\nu$ for axes. Panels (a) and (c) present the calculated phase diagrams near $\nu=1/3$ and 1/5, respectively \cite{Zhao.PRL.2018}. The yellow and blue regions indicate the predicted WC and liquid phases. So far, there have been no theoretical calculations for the $\nu$ ranges shown in panels (b) and (d), where a close competition between FQHSs and WC states is observed in our samples.  The open and solid circles denote the liquid and WC phases, respectively. The half-filled circles denote developing FQHSs in close competition with WC phases. The red and black circles represent two estimates of $\kappa$ for our 2DHS, as discussed in the text and Fig. \ref{fig:LLanalysis}(c). }
  \label{fig:phase}
\end{figure*}

Figure \ref{fig:phase} shows WC quantum melting phase diagrams, with LLM parameter $\kappa$ and $\nu$ for axes. The solid lines in panels (a) and (c) represent calculated phase boundaries between the WC and liquid states from Ref. \cite{Zhao.PRL.2018}. The red and black circles represent two estimates of $\kappa$ for our 2DHS shown in Fig. \ref{fig:LLanalysis}. Numerical calculations \cite{Zhao.PRL.2018} indicate that, while the $\nu=$ 1/3 FQHS is very robust to LLM and survives even for $\kappa>18$ [see Fig. \ref{fig:phase}(a)], the $\nu=$ 1/5 FQHS gives way to a WC phase for $\kappa\gtrsim10$ [Fig. \ref{fig:phase}(c)]. This is consistent with experimental results \cite{Kevin.PRR.2021, Ma.PRL.2020}; also see Figs. \ref{fig:phase}(a) and (c). Although there are no existing theories for the role of LLM at even smaller fillings near $\nu=$ 1/7, based on the trend discussed above, one would expect that FQHSs are more fragile at large $\kappa$ and would lose the competition to WC phases. Therefore, our observation of numerous odd- and even-denominator FQHSs at extremely small fillings, e.g. at $\nu=$ 1/7 and 1/6, in the presence of severe LLM ($\kappa\simeq5$) is surprising and calls for new theories and numerical calculations.

Our data points in Figs. \ref{fig:phase}(a, c) agree well with the calculated phase diagrams. As shown in Fig. \ref{fig:phase}(a), we observe fully developed FQHSs at $\nu=1/3$ and 2/5 (open circles), and a reentrant insulating phase between $\nu=1/3$ and 2/5 (solid circles), consistent with the expectation of a LLM-induced WC phase flanked by robust FQHSs at $\nu=1/3$ and 2/5. 

For $1/5\lesssim \nu \lesssim 2/9$, we observe developing FQHSs at $\nu=1/5$ and 2/9 [half-filled circles in Fig. \ref{fig:phase}(c)], i.e., sharp $R_{xx}$ minima that rise with decreasing $T$, signaling the close competition between FQHSs and WC states. Such a competition is also revealed by the calculated phase diagram shown in Fig. \ref{fig:phase}(c): Based on our estimates of $\kappa$ for the 2DHS, theory predicts that FQHSs only win in a narrow range of $\nu$ near 1/5 and 2/9, while WC states prevail elsewhere. Therefore, in real systems like our 2DHS, a small variation of $\nu$ caused by a minuscule density inhomogeneity or disorder can lead to the formation of WC domains and prevent the percolation of the FQH liquid. This explains why in our data the $R_{xx}$ minima do not reach zero at $\nu=1/5$, 2/9 and also other fractional $\nu$ where FQHSs emerge in the WC-dominated regime.

For the $\nu$ ranges shown in Figs. \ref{fig:phase}(b) and (d), we observe a fully developed FQHS at $\nu=2/7$ and numerous developing FQHSs in close competition with WC states. However, no theoretical calculations exist for these $\nu$ ranges.

\subsection{\label{sec:level2}Comparison with GaAs and AlAs 2DESs}

Next, we compare our results with recent findings in ultrahigh-quality GaAs and AlAs 2DESs at extremely small $\nu<1/5$ \cite{Chung.PRL.2022, Singh.PRB.2024, Wang.Preprint.2024}. GaAs 2DESs are considered hallmark platforms in quantum Hall physics thanks to their record-high mobility \cite{Chung.NatMater.2021, Chung.PRB.2022}. In state-of-the-art GaAs 2DESs, developing Jain-sequence FQHSs of $^6$CFs are seen on the flanks of $\nu=1/6$, at $\nu=1/7$, 2/13, 3/19, 3/17, 2/11, and 1/5 \cite{Wang.Preprint.2024}. A sharp $R_{xx}$ minimum at 1/6, signaling the emergence of a developing FQHS, is seen only when the 2DES is confined to a relatively wide QW, and has a single-layer, albeit thick, charge distribution \cite{Wang.Preprint.2024}. Similar to the $\nu=1/6$ FQHS in our 2DHS, the $\nu=1/6$ FQHS in GaAs 2DESs likely arises from the pairing of $^6$CFs \cite{Jain.privatecommunication}. Rather than LLM, it is the substantial electron layer thickness that softens the electron-electron Coulomb repulsion and facilitates CF pairing.

AlAs 2DESs also experience severe LLM at small $\nu$ thanks to their large effective mass and small dielectric constant compared to GaAs 2DESs. In AlAs 2DESs, signatures of developing FQHSs were very recently observed at $\nu=1/7$ and 2/11 with $\kappa\simeq4$ \cite{Singh.PRB.2024}. These findings qualitatively align with our data, although our 2DHS also shows hints of more high-order Jain-sequence FQHSs, e.g., at $\nu=3/17$ and 2/13. Unlike the emerging FQHSs at $\nu=1/4$ and 1/6 in our 2DHSs, AlAs 2DESs do not show any FQHS feature at these even-denominator fillings. At first sight, the absence of $\nu=1/4$ and 1/6 FQHSs in AlAs might appear to contradict our interpretation that these states arise from the pairing of $^4$CFs and $^6$CFs induced by LLM. However, several differences between the systems should be noted: (i) AlAs 2DESs have highly anisotropic Fermi contours, whereas GaAs 2DHSs at low densities have nearly isotropic Fermi contours. (ii) As discussed earlier, the LLs of GaAs 2DHSs are complex and nonlinear; on the other hand, the AlAs 2DESs have simple LLs. (iii) The quality of GaAs 2DHSs is generally superior to that of AlAs 2DESs, as reflected by the order-of-magnitude larger mobility of the GaAs 2DHSs \cite{Chung.PRM.2022, Gupta.PRM.2024} compared to AlAs 2DESs \cite{Chung.PRM.2018}.

\subsection{\label{sec:level2}FQHSs vs CF Fermi seas at $\nu=$ 1/4 and 1/6}

\begin{figure}[t!]
  \begin{center}
    \psfig{file=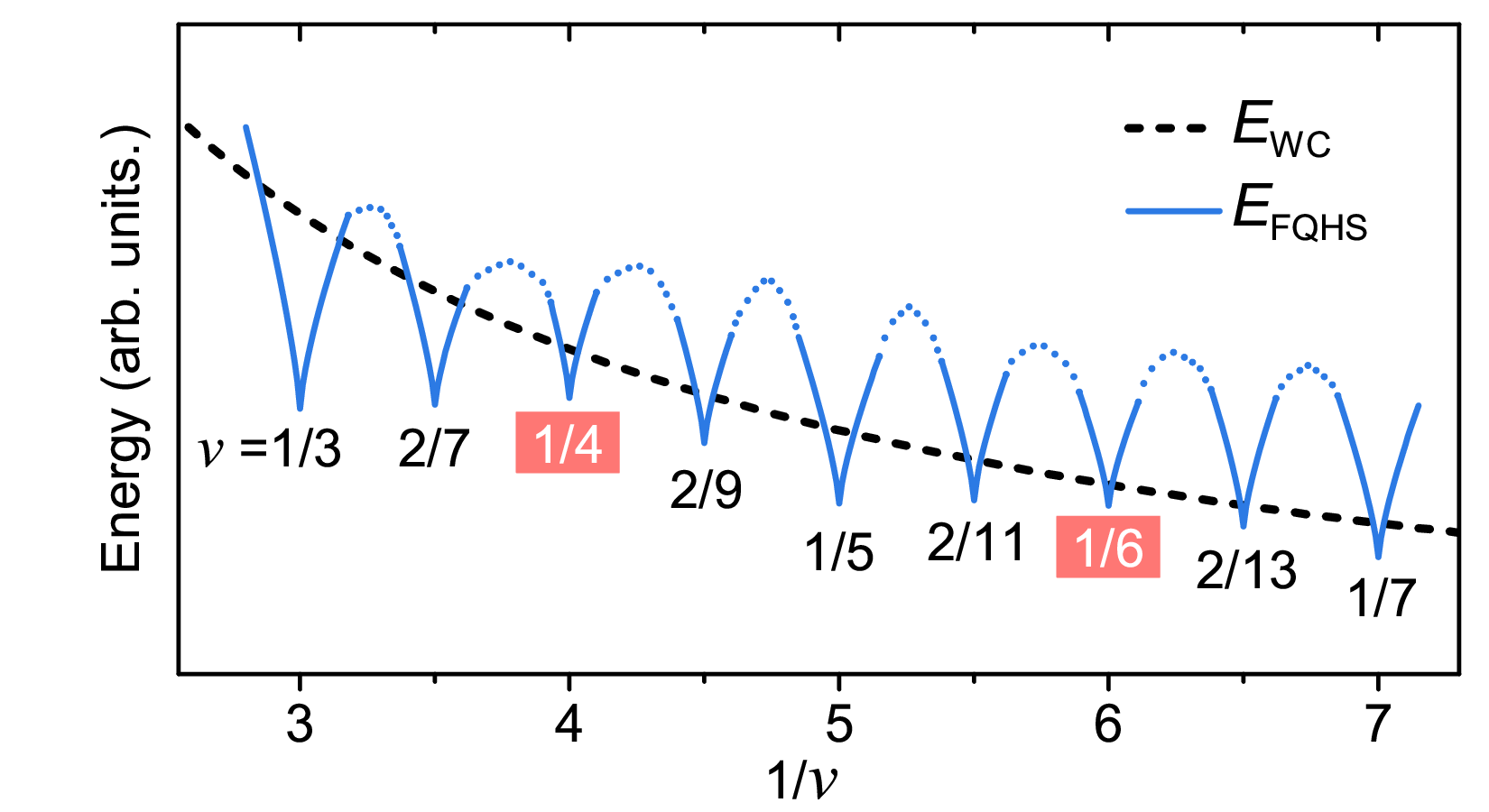, width=0.48 \textwidth}
  \end{center}
  \caption{\label{schematic} Schematic diagram showing the energies of WC and FQHSs vs $1/\nu$ for $\nu\lesssim 1/3$. While the energy of WC is expected to be a smooth function of $1/\nu$ \cite{Shayegan.review.2006, Archer.PRL.2013}, the energies of FQHSs exhibit downward cusps at proper fractional $\nu$, and can dip below the WC energy so that the WC could be reentrant on the flanks of a FQHS. This explains our observation of (developing) FQHSs at odd- and even-denominator fillings flanked by WC states (Fig. \ref{fig:Tdep}). For clarity, we only show the energies of FQHSs at $\nu=1/n$ and $2/n$, where $n$ is an integer. }
  \label{fig:schematic}
\end{figure}

Another potentially competitive ground state at $\nu=$1/6 (and 1/4) is a metallic Fermi sea of $^6$CFs ($^4$CF) at zero \textit{effective} magnetic field. One might take the $R_{xx}$ minima at $\nu=1/6$ and 1/4 as evidence for a CF Fermi sea at these fillings, similar to what is observed near $\nu=1/2$. While such an interpretation is also novel as there has been no experimental or theoretical evidence for such a phase deep in the low-filling, insulating, WC regime, we argue that it is unlikely that the $R_{xx}$ minima we observe at $\nu=1/6$ and 1/4 are indications of Fermi seas of $^4$CFs and $^6$CFs, but rather strongly favor developing FQHSs. First, our temperature dependence data reveal that the $R_{xx}$ minima at $\nu=1/4$ and 1/6 highly resemble those observed at surrounding odd-denominator, Jain-sequence fillings where developing FQHSs are expected. Second, theoretical calculations suggest that substantial LLM can induce $^4$CF pairing at $\nu=1/4$, and thereby favor a FQHS over a CF Fermi sea \cite{Zhao.PRL.2023}. Similar physics may be extended to $\nu=1/6$. Third, the sharp $R_{xx}$ minimum observed in the WC regime at $\nu=1/6$ (and 1/4) indicates that the $\nu=1/6$ (and 1/4) state is flanked by WC states. This strongly favors the interpretation of FQHS over $^6$CF ($^4$CF) Fermi sea: The energies of FQHSs show downward cusps as a function of $\nu$, and can dip below the WC energy (see Fig. 7) so that the WC could be reentrant on the flanks of a FQHS \cite{Archer.PRL.2013, Jiang.PRL.1990, Zuo.PRB.2020}. On the other hand, the energies of both CF Fermi sea and WC states are smooth functions of $\nu$, and thus one would expect only a single transition between these two states. It is also worth noting that the observation of Jain-sequence FQHSs around a given even-denominator filling does not necessarily indicate the presence of a CF Fermi sea at that filling. Indeed, a FQHS at $\nu=1/2$ flanked by numerous Jain-sequence FQHSs is well established in 2DESs confined to wide GaAs QWs \cite{Suen.PRL.1992, Singh.NatPhys.2024}.

\section{\label{sec:level1}Summary}

Our observation of numerous developing FQHSs at extremely small, even- and odd-denominator $\nu$ competing with WC states highlights the rich physics in the lowest LL of ultrahigh-quality GaAs 2DHSs. The robustness of FQHSs in the presence of LLM is surprising, and should inspire further theoretical calculations. Also puzzling are the anomalously large activation energies we measure in the WC regime (Fig. \ref{fig:activation}); they suggest that the WC intrinsic defects may be different in a 2D system with severe LLM.

\begin{acknowledgments}
We acknowledge support by the National Science Foundation (NSF) (Grant No. DMR 2104771) for measurements, the U.S. Department of Energy (DOE) Basic Energy Sciences (Grant No. DEFG02-00-ER45841) for sample characterization, and the National Science Foundation (NSF) (Grant No. ECCS 1906253), the Eric and Wendy Schmidt Transformative Technology Fund, and the Gordon and Betty Moore Foundation’s EPiQS Initiative (Grant No. GBMF9615 to L.N.P.) for sample fabrication. Our measurements were partly performed at the National High Magnetic Field Laboratory (NHMFL), which is supported by the NSF Cooperative Agreement No. DMR 2128556, by the State of Florida, and by the DOE. This research is funded in part by QuantEmX grants from Institute for Complex Adaptive Matter and the Gordon and Betty Moore Foundation through Grant No. GBMF9616 to C. W. and S. K. S. We thank A. Bangura, R. Nowell, G. Jones, and T. Murphy at NHMFL for technical assistance, and P. T. Madathil, A. C. Balram and J. K. Jain for illuminating discussions.
\end{acknowledgments}

\appendix

\begin{figure}[t!]
  \begin{center}
    \psfig{file=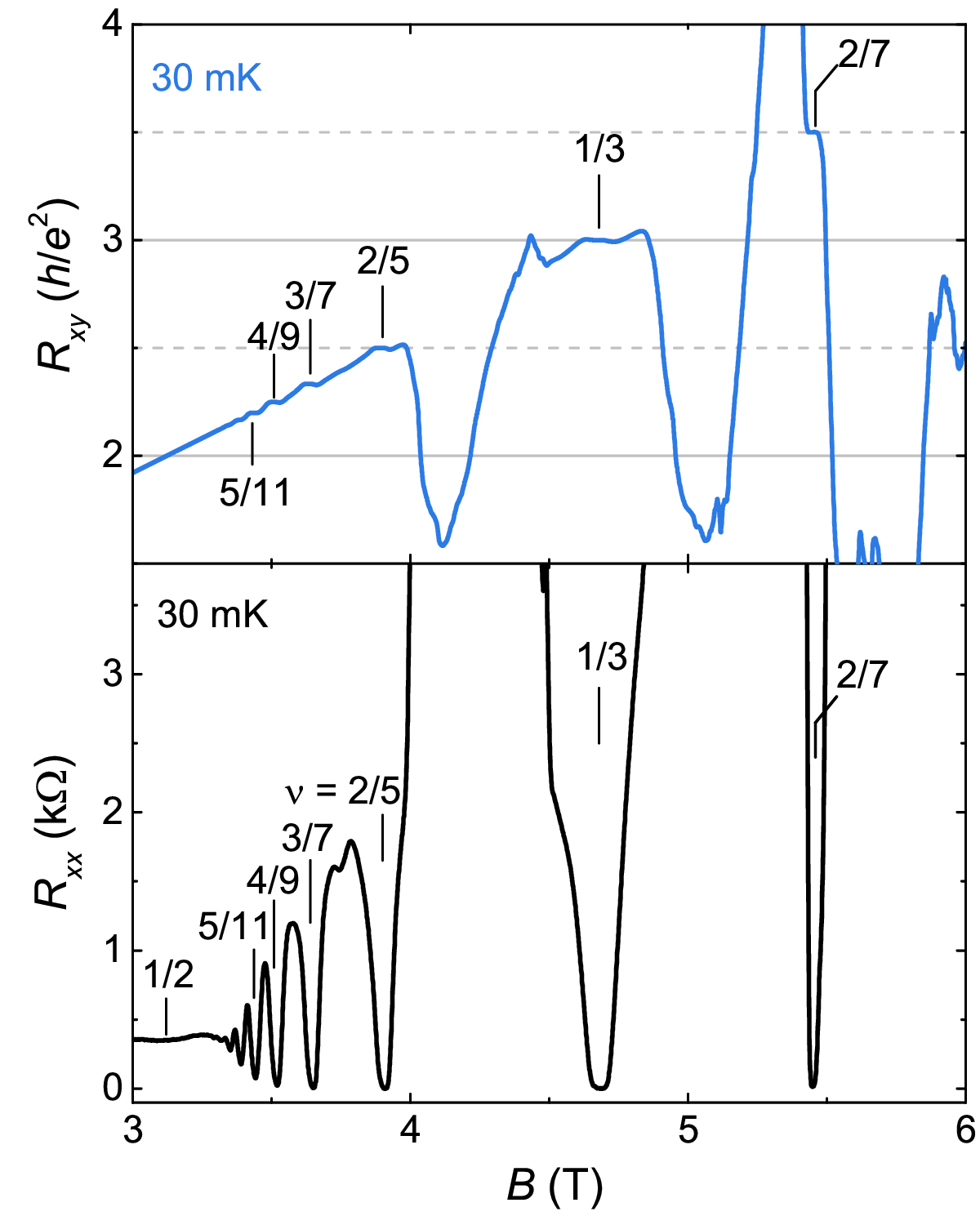, width=0.45 \textwidth}
  \end{center}
  \caption{\label{Hall}
   {\bf Hall data.} (a) Longitudinal resistance ($R_{xx}$) and (b) Hall resistance ($R_{xy}$) vs $B$ traces measured with $I=100$ nA at $T\simeq30$ mK. Different LL fillings are marked by vertical lines.
   }
  \label{fig:Hall}
\end{figure}

\begin{figure*}[t]
  \begin{center}
    \psfig{file=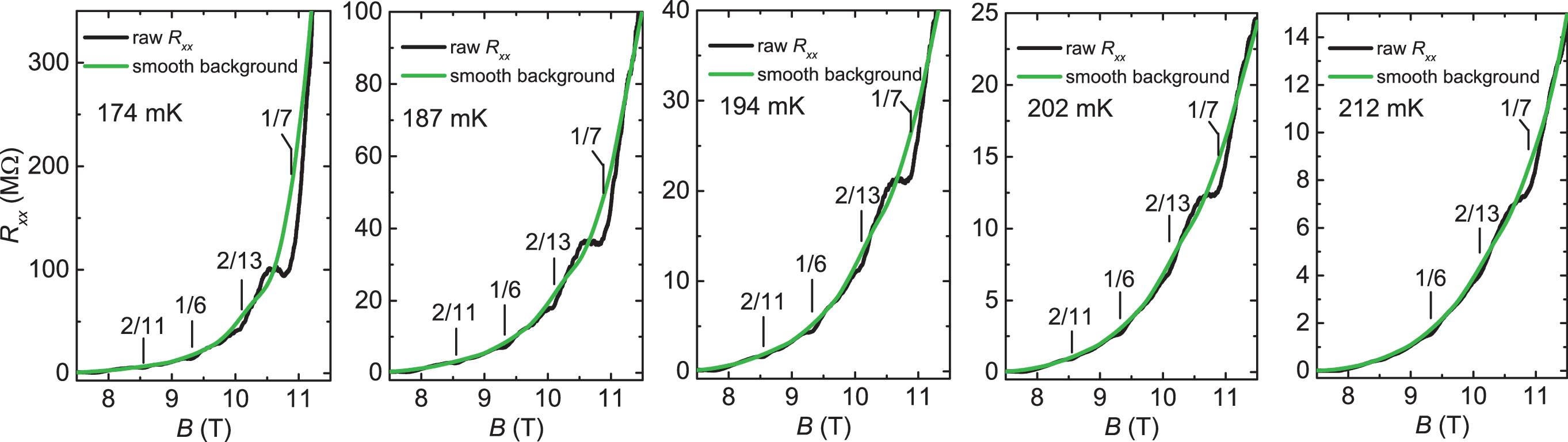, width=1 \textwidth}
  \end{center}
  \caption{\label{background}
   {\bf Resistance background subtraction.} Raw $R_{xx}$ vs $B$ traces (in black) and their corresponding smooth background ($R_{bg}$, in green) at different temperatures. $R_{xx}$ clearly dips below $R_{bg}$ at $\nu=2/11$, 1/6, 2/13, and 1/7.
   }
  \label{fig:background}
\end{figure*}

\begin{figure*}[t]
  \begin{center}
    \psfig{file=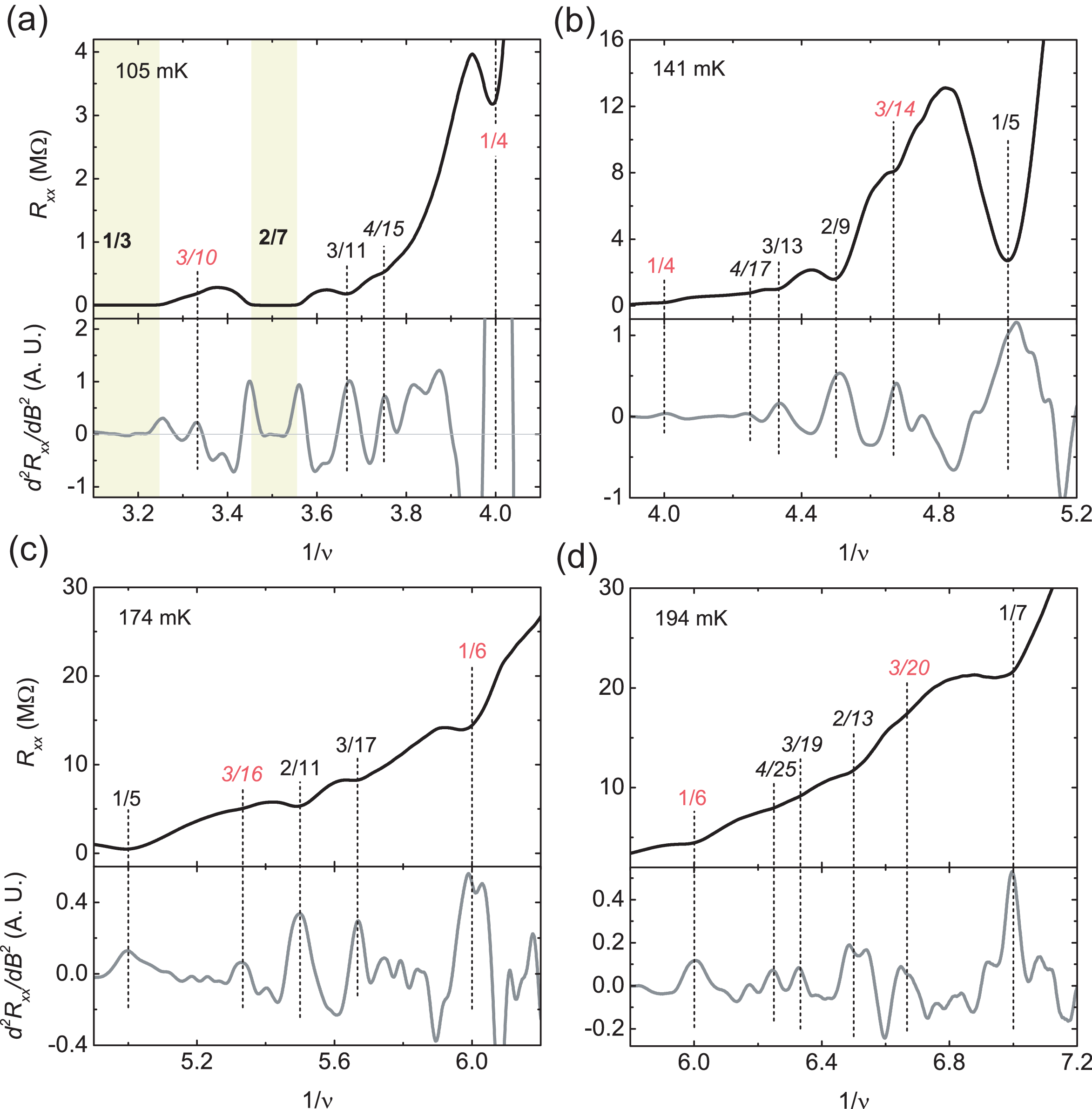, width=0.8 \textwidth}
  \end{center}
  \caption{\label{derivative}
   {\bf Evidence for developing FQHSs.} $R_{xx}$ and $d^2R_{xx}/dB^2$ for our 2DHS with density $3.8\times10^{10}$ cm$^{-2}$ are plotted as a function of inverse filling $1/\nu$ for different temperatures and ranges of $1/\nu$. Vertical dotted lines indicate the positions where developing FQHSs are observed, evinced by $R_{xx}$ minima (or inflection points), and by sharp peaks in $d^2R_{xx}/dB^2$. 
   }
  \label{fig:derivative}
\end{figure*}

\section{Hall data}

Figure \ref{fig:Hall}(b) shows Hall resistance ($R_{xy}$) as a function of $B$ for the 2DHS discussed in the main text. In Fig. \ref{fig:Hall}(a), we present $R_{xx}$ data for the same range of $B$. We observe FQHSs at $\nu=1/3$, 2/5, 3/7, ..., and also at $\nu=2/7$, evinced by deep $R_{xx}$ minima and quantized Hall plateaus. We note that in the ranges $\nu<2/7$, $1/3>\nu>2/7$, and $2/5>\nu>1/3$, where the 2DHS is highly insulating, the $R_{xy}$ trace significantly deviates from the linear trend because of the effect of $R_{xx}$ mixing. This is a well-known problem, and manifests itself in the low-$\nu$ regime where there is a close competition between WC states and FQHSs \cite{Goldman.PRL.1988, Sajoto.PRL.1993, Goldman.PRL.1993}. Although flanked by insulating phases, the FQHSs at $\nu=1/3$ and 2/7 are fully developed at $T\simeq30$ mK, with $R_{xx}$ reaching zero and flat $R_{xy}$ plateaus quantized at $3h/e^2$ and $7h/2e^2$, respectively. 

\begin{table*}[t]
\caption{\label{tab:table} List of FQHSs observed in our 2DHS at $\nu\leq1/3$. States denoted in boldface are fully developed FQHSs, well established by the observation of vanishing $R_{xx}$ and quantized $R_{xy}$ plateaus. States represented in regular font indicate developing FQHSs, evinced by discernible $R_{xx}$ minima on top of an insulating background. States shown in italic denote observed inflection points in $R_{xx}$. States at even-denominator fillings are highlighted in red.}
\begin{ruledtabular}
\begin{tabular}{ccccccc}
 CF $\Lambda$L fillings&1&3/2&2&3&4&$\infty$\\ \hline
 \multirow{2}{*}{$^4$CF}&$\boldsymbol{1/3}$&\textcolor{red}{$\mathit{3/10}$}&$\boldsymbol{2/7}$&3/11&$\mathit{4/15}$&\textcolor{red}{1/4}\\ &1/5&\textcolor{red}{$\mathit{3/14}$}&2/9&3/13&$\mathit{4/17}$&\textcolor{red}{1/4}\\ \hline
 \multirow{2}{*}{$^6$CF}&1/5&\textcolor{red}{$\mathit{3/16}$}&2/11&3/17& &\textcolor{red}{1/6}\\ &1/7&\textcolor{red}{$\mathit{3/20}$}&$\mathit{2/13}$&$\mathit{3/19}$&$\mathit{4/25}$&\textcolor{red}{1/6}\\
\end{tabular}
\end{ruledtabular}
\end{table*}

\section{Resistance background subtraction and $\boldmath{d^2R_{xx}/dB^2}$ data at small $\nu$}
In order to better visualize the FQHS features in $R_{xx}$ amidst the dominant insulating behavior caused by the competing WC states, we extract the oscillatory component ($\Delta R_{xx}$) of the $R_{xx}$ data. Here $\Delta R_{xx}=R_{xx}-R_{bg}$ is defined as the resistance after subtracting a smooth, increasing background $R_{bg}$. As we show in Fig. \ref{fig:deltaR}, sharp minima are revealed in the $\Delta R_{xx}/R_{bg}$ vs $B$ traces. Figure \ref{fig:background} displays the raw $R_{xx}$ vs $B$ traces (black) and their corresponding $R_{bg}$ (green) at different temperatures. The resistance background $R_{bg}$, contributed by the insulating phase, is estimated using a Savitzky-Golay filter with a 1 T window applied to $R_{xx}$ vs $B$ data. This window size captures the main trend of the slowly varying part of $R_{xx}$ while minimizing artifacts.

Another effective method to extract FQHS features from highly insulating $R_{xx}$ data is to calculate the second derivative, $d^2R_{xx}/dB^2$. Figure \ref{fig:derivative} shows $R_{xx}$ (top panels) and $d^2R_{xx}/dB^2$ (bottom panels) as a function of inverse filling $1/\nu$ for different temperatures and ranges of $1/\nu$. At $\nu=1/3$ and 2/7, FQHSs are fully developed, evinced by vanishing $R_{xx}$ spanning a range of $1/\nu$ and zero $d^2R_{xx}/dB^2$ flanked by two peaks; see yellow-shaded part in Fig. \ref{fig:derivative}(a). At small $\nu$ deep in the insulating regime where WC states are dominant, a developing FQHS manifests itself as a minimum or an inflection point in $R_{xx}$ with a maximal curvature and therefore exhibits a peak in $d^2R_{xx}/dB^2$. In Fig. \ref{fig:derivative}, we observe clear $R_{xx}$ minima superimposed on a highly insulating background at odd-denominator fillings $\nu=3/11$, 3/13, 2/9, 1/5, 2/11, 3/17, and 1/7, signaling the developing Jain-sequence states of four-flux and six-flux composite fermions ($^4$CFs and $^6$CFs). Clear $R_{xx}$ minima are also seen at even-denominator fillings $\nu=1/4$ and 1/6, suggesting developing FQHSs, likely emerging from the pairing instability of $^4$CFs and $^6$CFs. In addition, we also see inflection points at certain fillings, including odd-denominator fillings, consistent with high-order Jain-sequence states of $^4$CFs ($\nu=4/15$ and 4/17) and $^6$CFs ($\nu=2/13$, 3/19, and 4/25), and even-denominator fillings at $\nu=3/10$, 3/14, 3/16, and 3/20.

\section{Summary of developing FQHSs at extremely small $\nu$}

In Table. I, we list all the filling factors $\nu$ at which FQHS features in $R_{xx}$ and/or $R_{xy}$ are observed in our 2DHS. States denoted in boldface are fully developed FQHSs, well established by the observation of vanishing $R_{xx}$ and quantized $R_{xy}$ plateaus. States represented in regular font indicate developing FQHSs, evinced by discernible $R_{xx}$ minima on top of an insulating background. States shown in italic denote observed inflection points in $R_{xx}$. States at even-denominator fillings are highlighted in red. We map these hole LL fillings $\nu$ to CF Lambda level ($\Lambda$L) fillings \cite{Jain.Book.2007} of $^4$CFs and $^6$CFs. The odd-denominator, Jain-sequence FQHSs can be mapped to interger QHSs of $^4$CFs and $^6$CF. Note that the states at even-denominator $\nu=3/10$, 3/14, 3/16, and 3/20 can all be mapped to a CF $\Lambda$L filling of 3/2. These states likely emerge from the same mechanism as those observed at $\nu=3/4$ and 3/8, which can be interpreted as "next-generation" even-denominator FQHSs of CFs with $\Lambda$L filling of 3/2 for two-flux CFs. \cite{Wang.PRL.2022, Wang.PNAS.2023}.

\section{Magneto-transport data for a different sample}

\begin{figure*}[t]
  \begin{center}
    \psfig{file=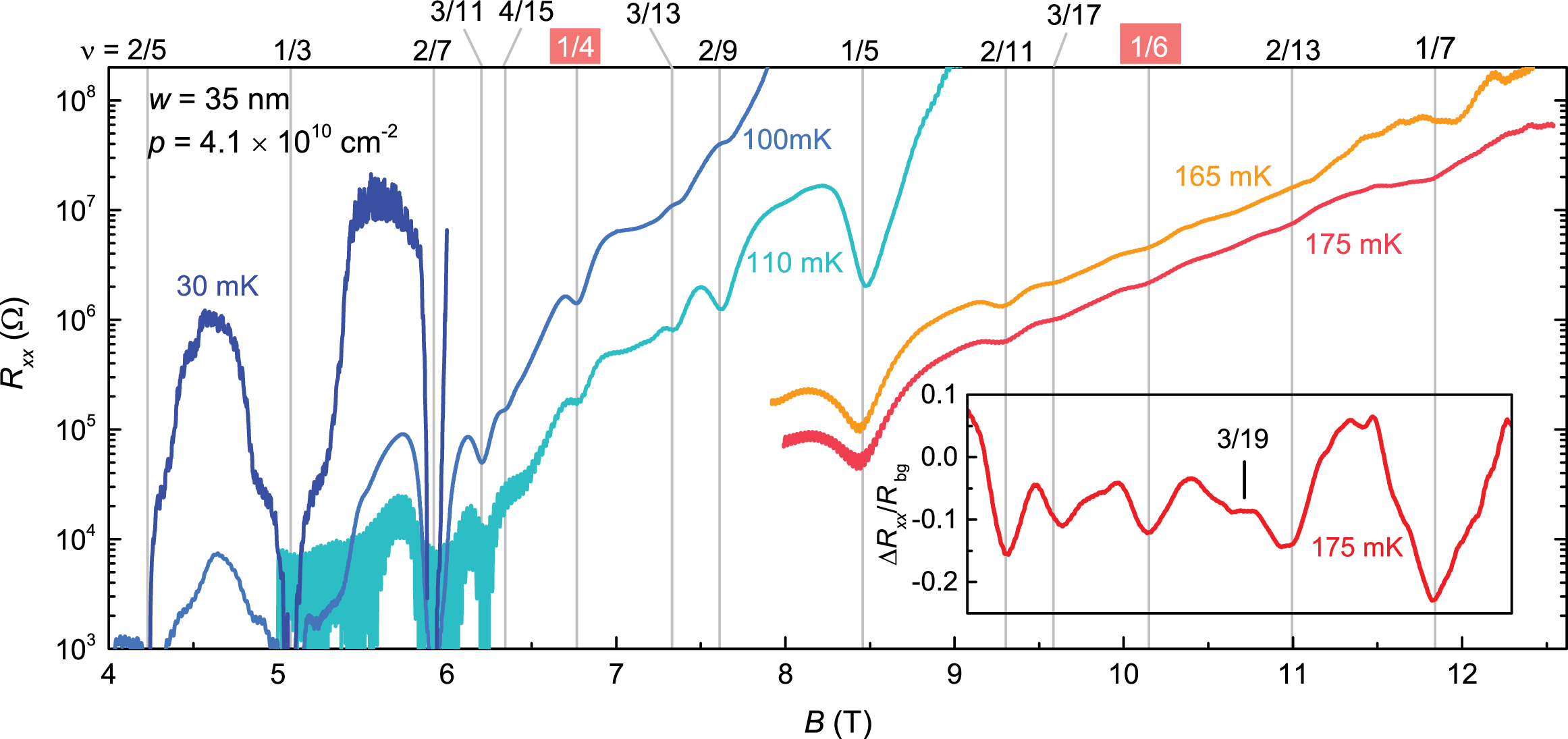, width=1 \textwidth}
  \end{center}
  \caption{\label{sampleB}
   {\bf Magneto-transport data for a 35-nm-wide quantum well.} $R_{xx}$ vs $B$ traces measured with $I=0.1$ nA at different temperatures. The vertical lines mark the $B$ positions of fractional filling factors indicated on the top axis. Inset presents $\Delta R_{xx}/R_{bg}$ in the range $9\lesssim B\lesssim12.3$ T for $T\simeq175$ mK. }
  \label{fig:sampleB}
\end{figure*}

In Fig. \ref{fig:sampleB}, we present $R_{xx}$ data for a different sample measured at different temperatures. The 2DHS is confined to a 35-nm-wide quantum well, and has a hole density of $4.1\times10^{10}$ cm$^{-2}$. We observe qualitatively similar behaviors compared to the sample we discuss in the main text (see Fig. 2 of the main text): (i) The 2DHS shows an insulating behavior in the filling ranges $2/5>\nu>1/3$, $1/3>\nu>2/7$, and $\nu<2/7$. (ii) Fully developed FQHSs are seen at $\nu=1/3$ and 2/7, evinced by vanishing $R_{xx}$. (iii) $R_{xx}$ minima and inflection points superimposed on a highly insulating background are observed at odd-denominator fillings $\nu=n/(4n\pm1)$ and $\nu=n/(6n\pm1)$, where $n$ are integers. These features are consistent with devloping Jain-sequence FQHSs of $^4$CFs and $^6$CFs. (iv) At even-denominator fillings $\nu=1/4$ and 1/6, respectively, a $R_{xx}$ minimum and an inflection point are observed, suggesting developing FQHSs. The $R_{xx}$ minima and inflection points at $\nu=2/11$, 3/17, 1/6, 2/13, and 1/7 are clearly revealed as sharp minima in $\Delta R_{xx}/R_{bg}$ in the inset of Fig. \ref{fig:sampleB}, where we present $\Delta R_{xx}/R_{bg}$ vs $B$ at 175 mK. These observations demonstrate that the competing FQHSs and WC phases at small LL fillings we report in the main text are robust and reproducible phenomena in ultrahigh-quality 2DHSs.

\section{Landau level calculations}

We performed LL calculations using the multi-band envelope function approximation based on the $8\times8$ Kane Hamiltonian \cite{Winkler.Book.2003} for our 2DHS studied in the main text. The model also uses the axial approximation which neglects the effect of cubic anisotropy of the crystal structure. Figure \ref{fig:LLs}(a) shows the calculated $E$ vs $B$ LL diagram. The LLs are nonlinear as a function of $B$ and are not evenly spaced in $E$ at a particular $B$ because of the strong coupling between heavy-hole (HH) and light-hole (LH) subbands \cite{Winkler.Book.2003}. This implies that the cyclotron energy may not be well defined for GaAs 2DHSs.

\begin{figure*}[t]
  \begin{center}
    \psfig{file=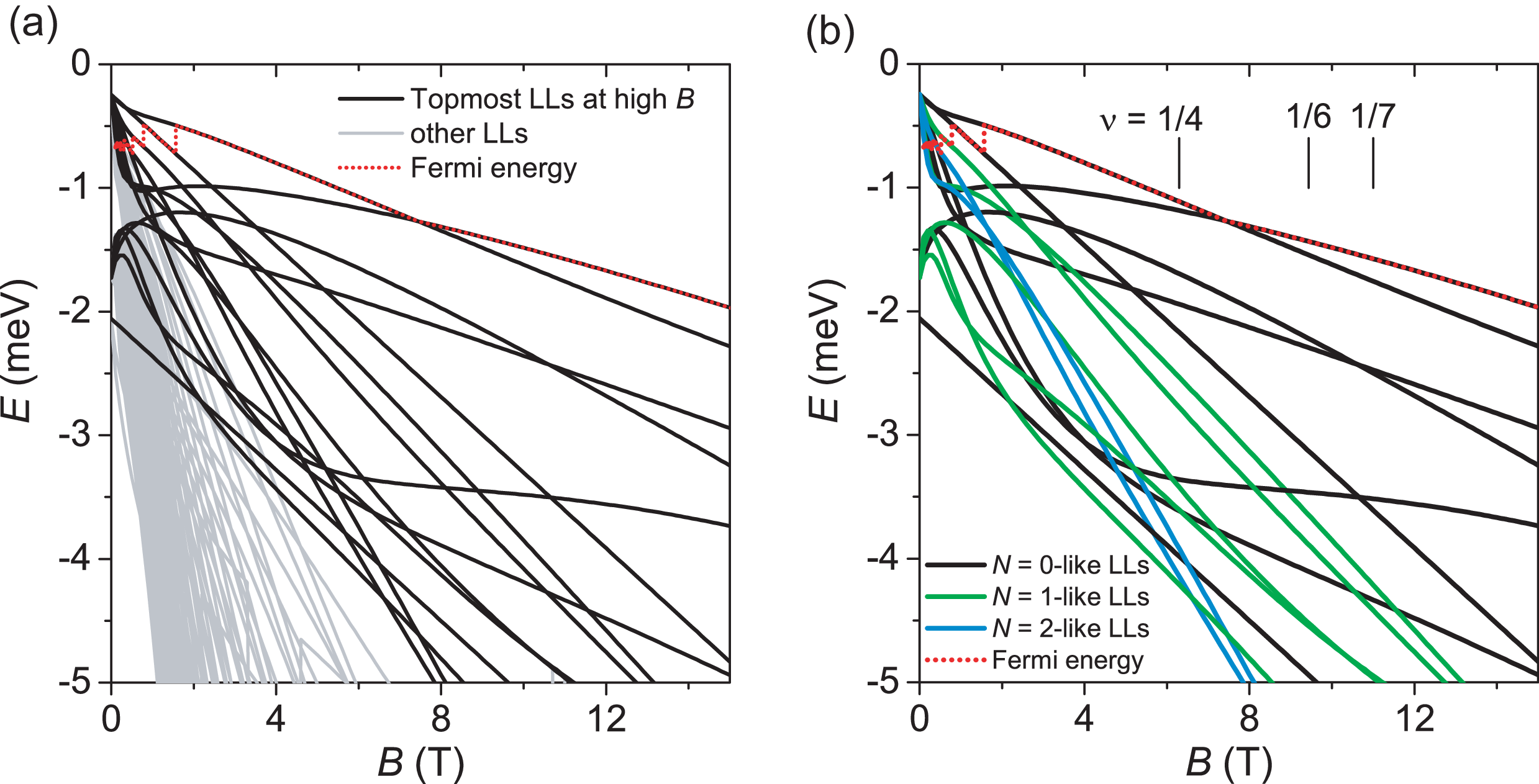, width=0.8\textwidth}
  \end{center}
  \caption{\label{fig:LLs}
    {\bf Landau level diagrams.} (a) Calculated $E$ vs $B$ Landau level diagram. The red dotted line traces the Fermi energy $E_F$. (b) Topmost LLs at high $B$ are color-coded according to their dominant orbital component in the field range $4<B<12$ T where competing FQHSs and WC states are observed. $B$ positions of typical fillings are marked with vertical lines.
    }
  \label{fig:LLs}
\end{figure*}

\begin{figure*}[t]
  \begin{center}
    \psfig{file=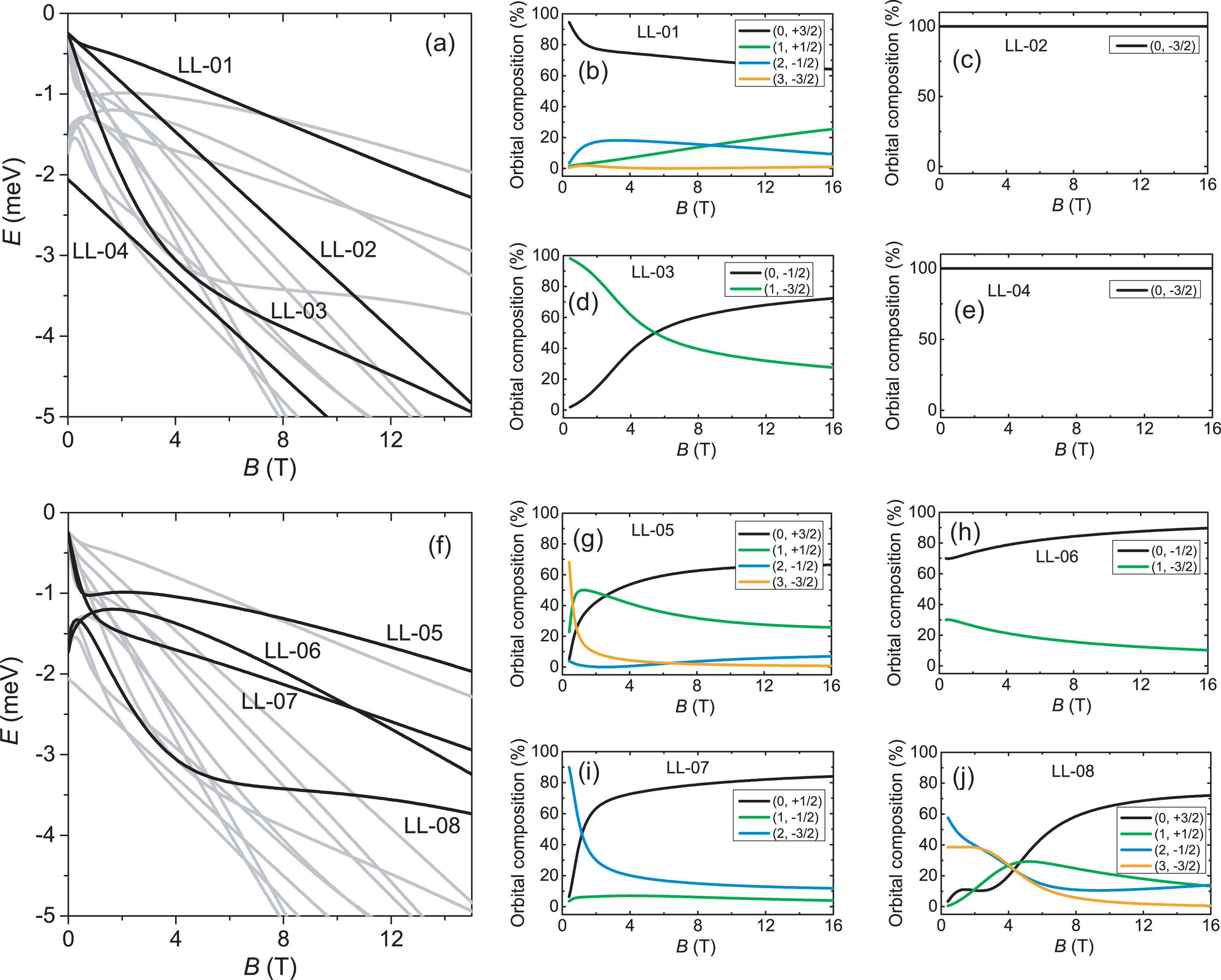, width=1\textwidth}
  \end{center}
  \caption{\label{fig:LLN0}
    {\bf $\boldsymbol{N=0}$-like Landau levels and their orbital compositions.} (a) $E$ vs $B$ LL diagram with LL-01, LL-02, LL-03, and LL-04 highlighted. (b-e) Orbital compositions for LL-01, LL-02, LL-03, and LL-04, respectively. The Landau orbital quantum number $N$ and spin quantum number $s_z$ are shown in the legend boxes as ($N$, $s_z$). (f) $E$ vs $B$ LL diagram with LL-05, LL-06, LL-07, and LL-08 highlighted. (g-j) Orbital compositions for LL-05, LL-06, LL-07, and LL-08, respectively.
    }
  \label{fig:LLN0}
\end{figure*}

\begin{figure*}[t]
  \begin{center}
    \psfig{file=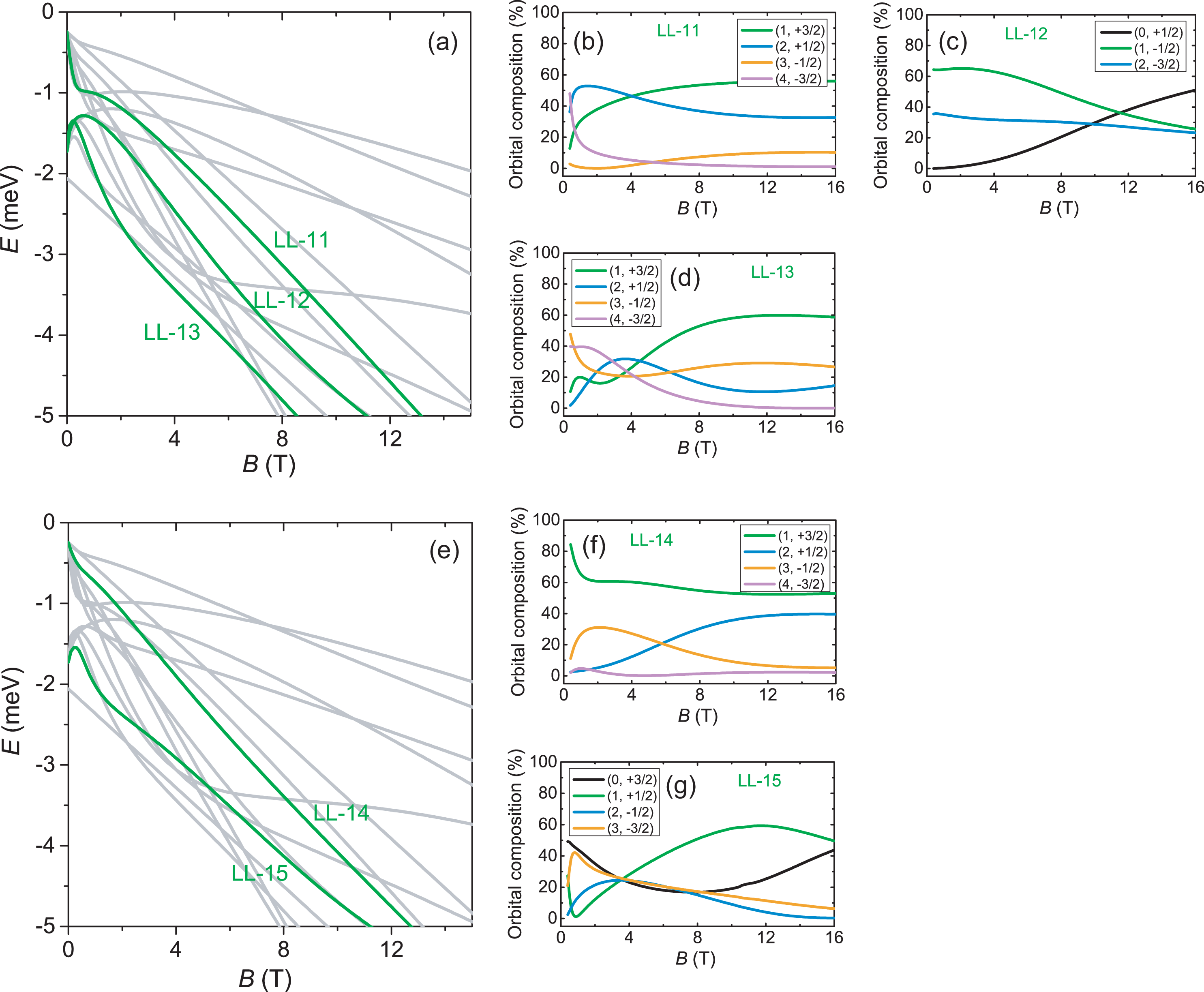, width=1\textwidth}
  \end{center}
  \caption{\label{fig:LLN1}
    {\bf $\boldsymbol{N=1}$-like Landau levels and their orbital compositions.} (a) $E$ vs $B$ LL diagram with LL-11, LL-12, and LL-13 highlighted. (b-d) Orbital compositions for LL-11, LL-12, and LL-13, respectively. (e) $E$ vs $B$ LL diagram with LL-14 and LL-15 highlighted. (f, g) Orbital compositions for LL-14 and LL-15, respectively.
    }
  \label{fig:LLN1}
\end{figure*}

\begin{figure*}[t]
  \begin{center}
    \psfig{file=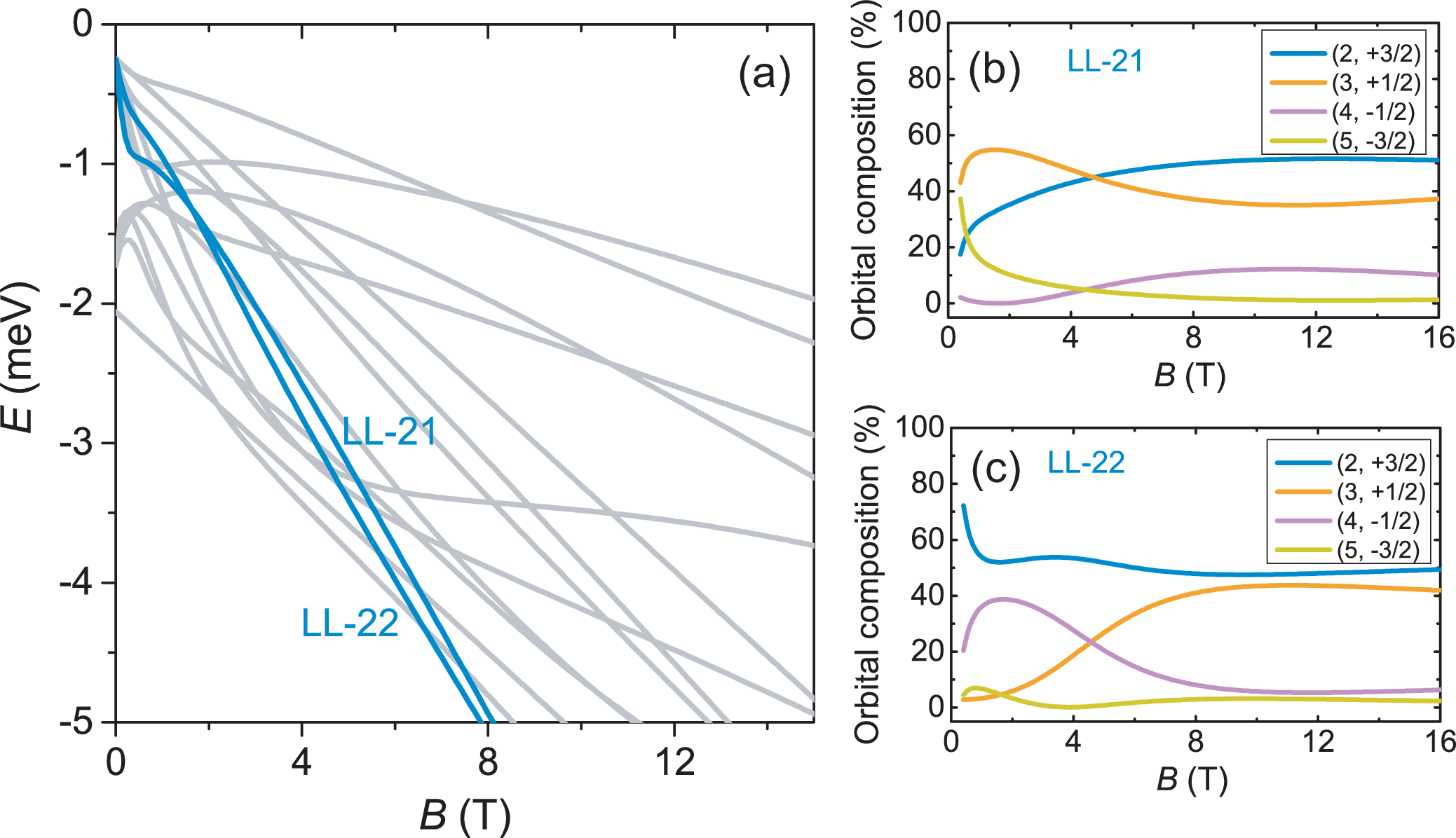, width=0.75 \textwidth}
  \end{center}
  \caption{\label{fig:LLN2}
    {\bf $\boldsymbol{N=2}$-like Landau levels and their orbital components.} (a) $E$ vs $B$ LL diagram with LL-21 and LL-22 highlighted. (b, c) Orbital compositions for LL-21 and LL-22, respectively. 
    }
  \label{fig:LLN2}
\end{figure*}

The nonlinearity as a function of $B$ is closely related to the fact that the hole LLs also have a more complex nature compared to electron LLs for a GaAs 2D electron system (2DES). Unlike the simple LLs for a GaAs 2DES that can be described by an orbital quantum number $N=0,1,2, \ldots$ and a spin quantum number $s=\pm1/2$, for hole LLs $N$ and $s$ are not good quantum numbers. While the eigenstates at zero $B$ (and in-plane wave vector $k_\|=0$) are purely HH with $s=\pm3/2$ or LH with $s=\pm1/2$, the HH-LH coupling at $B > 0$ leads to eigenstates with mixed HH and LH components.  Strictly speaking, the eigenstates in the accurate $8 \times 8$ Kane model have eight spinor components.  However, for the hole states relevant here the weights of the additional four spinor components are small relative to the HH and LH components, and they are therefore ignored in the following discussion.

We emphasize that the coupling and mixing of HH and LH components in a single LL is fundamentally different from the LL mixing (LLM) we discuss in the main text. The former is a consequence of HH-LH coupling in the single-particle picture (the single-particle eigenstates in a quantizing magnetic field are already a superposition of Landau harmonic oscillators), while the latter (LLM) is a many-body effect induced by the Coulomb interaction. LLM happens on top of what is already present due to HH-LH coupling. As we discuss next, in order to make quantitative estimates of the LLM parameter $\kappa$, we simplify the problem by expanding the hole LLs in a basis of four spinors and treating each hole LL as a single component LL described by its leading spinor component.

Here we limit our analysis to the topmost LLs at high $B$ [shown in Fig. \ref{fig:LLs}(a) in black] which are most relevant to our study of correlated states at high $B$, and other LLs [shown in Fig. \ref{fig:LLs}(a) in grey] that are further away from the Fermi energy are neglected for simplicity and clarity. In Fig. \ref{fig:LLs}(b), we replot Fig. 4(a) of the main text, showing the most relevant LLs, color-coded according to their dominant orbital component in the field range $4<B<12$ T where competing FQHSs and WC states are observed.

Next, we discuss in detail how we determine the dominant orbital component for the LLs shown in Fig. \ref{fig:LLs}(b). The LLs at $B > 0$ can be decomposed into a basis of four spinors, writen as:
\begin{equation}
  \label{eq:ax-state}
  \psi_\mathcal{N} = \sum_s \ket{N = \mathcal{N} - s - \frac{3}{2}}
  \, \xi^\mathcal{N}_s u_s
\end{equation}
with $N\geq0$. Here $u_s$ are the base spinors with spin $s$ and $\xi^\mathcal{N}_s$ give the relative weights of the HH and LH spinors. $\mathcal{N} = 0, 1, 2, \ldots$ are quantum numbers that label the energies of the LLs \cite{Winkler.Book.2003}. In Eq. (E1), a LL described by a particular quantum number $\mathcal{N}$ can be decomposed into a basis of four spinors $u_s$, each of which can be described by a spin quantum number $s=\pm1/2$ or $\pm3/2$ and is coupled to a Landau harmonic oscillator with orbital quantum number $N=\mathcal{N}-s-\frac{3}{2}$. The relative weight $\xi^\mathcal{N}_s$ of each spinor for a LL is equivalent to the relative weight of the $N=\mathcal{N}-s-\frac{3}{2}$ orbital component for this LL, providing information for LL orbital compositions. In Figs. 13-15, we present calculated orbital compositions for $N=0$-like, $N=1$-like, and $N=2$-like LLs shown in Fig. \ref{fig:LLs}(b), respectively.

Some LLs are rather simple in their orbital compositions. For example, LL-02 is a purely $N=0$, $s_z=-3/2$ LL [$100\%$ (0, -3/2) component; see Fig. \ref{fig:LLN0}(c)] \cite{Broido.PRB.1985}. It has a linear $E$ vs $B$ dependence [see Fig. \ref{fig:LLN0}(a)], reminiscent of LLs for GaAs 2DESs. LL-02 is decoupled from all other LLs, because it has a quantum number $\mathcal{N}=0$ and the only combination of $N$ and $s_z$ that meets the condition $\mathcal{N}=N+s+\frac{3}{2}$ is $N=0$ and $s_z=-3/2$, denoted as (0, -3/2). Other LLs, however, are more complex and have up to four orbital components whose weights vary with $B$, leading to a highly nonlinear $E$ vs $B$ dependence. For some LLs, the leading orbital component even switches at finite $B$ as a result of LL anti-crossings. Taking LL-03 as an example, it has two orbital components (0, -1/2) and (1, -3/2). While the (1, -3/2) component dominates at low $B$, its weight decreases monotonically from $100\%$ to $25\%$ when $B$ increases from 0 to 20 T, and as a result, the (0, -1/2) component becomes dominant at high $B$. A crossover from the $N=1$-dominant regime to the $N=0$-dominant regime for LL-03 occurs at $\simeq5$ T, consistent with a significant change in slope at $\simeq5$ T in Fig. \ref{fig:LLN0}(a). Considering that we are interested in the physics at high $B$ ($\geq5$ T), LL-03 is categorized as a $N=0$-like LL. Similar logic is applied to categorize the LLs shown in Fig. \ref{fig:LLs}(b).

As noted in Sec. IV.A, the ratio $E_{\text{Coul}}/\Delta E$ obtained from our calculated LL fan diagram provides a good estimate of the LLM parameter $\kappa$ for our 2DHS. One may wonder whether this approach could offer insights into $\kappa$ for various GaAs 2DHSs with different sample parameters. Our preliminary calculations suggest that at certain $B$, the $\Delta E$ values remain approximately consistent across GaAs 2DHSs with varying quantum well widths and densities. This consistency implies that the plot of $E_{\text{Coul}}/\Delta E$ vs $B$ shown in Fig. \ref{fig:LLanalysis}(c) may serve as a reasonable estimate of $\kappa$ for GaAs 2DHSs with diverse sample parameters.


\begin{thebibliography}{99}
\bibitem{Footnote.deltaE} In GaAs 2DESs, where LLs are strictly linear with $B$, $\Delta E$ is simply equal to $E_{\text{cyc}}$. It is noteworthy that, for our 2DHS, between the topmost $N=0$-like and $N=1$-like LLs, there are multiple other $N=0$-like LLs whose leading spinor components have different spin quantum numbers $s_z=\pm3/2$ and $\pm1/2$. Some of these LLs originate from the second electric subband of the QW; see Appendix E for details of the spinor composition of these LLs. Here we ignore these LLs in our definition of $\Delta E$ because the separations between the lowest LL and these other $N=0$ LLs, in principle, do not affect the cyclotron energy. The effect of multiple $N=0$ LLs near the lowest LL is not negligible, and is not captured by $\kappa$. The role of the proximity of these $N=0$ LLs to the lowest LL is also important, but beyond the scope of this article.
\bibitem{Footnote.agreement} Note that both $E_{\text{Coul}}/E_{\text{cyc}}$ and $E_{\text{Coul}}/\Delta E$ here provide only rough estimations of $\kappa$. This is because hole LLs are not evenly spaced in $E$ as a consequence of their nonlinearity, and each hole LL is a superposition of four spinors with different $N$, as described in the text and Appendix E.
\end{thebibliography}
\end{document}